\documentclass{iopjournal}

\usepackage[english]{babel}
\usepackage{amsmath}
\usepackage{amssymb}
\usepackage{braket}
\usepackage{bm}
\usepackage{bbold}
\usepackage{booktabs}
\usepackage{tabularx}

\allowdisplaybreaks

\newcommand{\bfr}{\mathbf r}

\newcommand{\rmd}{\mathrm d}
\newcommand{\pvec}[1]{\vec{#1}\mkern2.5mu\vphantom{#1}}

\DeclareMathOperator\Erf{Erf}

\begin{document}

\articletype{Paper}

\title{Semi-regularised three-body pseudopotential for mean-field and 
beyond-mean-field calculations}


\author{V.~Guillon$^{1,\text{a}}$\orcid{0009-0006-1476-402X},
M.~Bender$^{1,\text{b}}$\orcid{0000-0001-8707-3410},
K.~Bennaceur$^{1,\text{c},*}$\orcid{0000-0002-6722-491X},
and Ph.~Da Costa$^{1,\text{d}}$\orcid{0009-0002-9776-616X}}


\affil{$^1$Universit{\'e} Lyon 1, CNRS, IP2I, UMR 5822, Villeurbanne, France \\ }
\affil{$^*$Author to whom any correspondence should be addressed.}


\email{$^\text{a}$v.guillon@ip2i.in2p3.fr, 
$^\text{b}$bender@ipnl.in2p3.fr,
$^\text{c}$bennaceur@ipnl.in2p3.fr,
$^\text{d}$p.da-costa@outlook.fr}




\begin{abstract}

We derive the most general form of a local leading-order 
semi-regularised three-body pseudopotential.
This particular form of pseudopotential is developed with 
the aim of generating contributions to the nuclear 
energy density functional (EDF) in both the particle-hole and 
particle-particle channels and, hence, to be usable in mean-field and 
beyond-mean-field calculations without ambiguities or mathematical difficulties. 
Once the EDF is obtained, analytical expressions of commonly considered
properties of infinite nuclear matter are provided.
Finally, the structure of the EDF and the associated mean fields are given for 
spherically-symmetric systems.

\end{abstract}


\section{Introduction\label{sec:introduction}}


Methods based on energy density functionals (EDFs) are widely-used tools to describe
a large variety of static and dynamic properties of atomic nuclei irrespective of 
their mass and asymmetry.
The starting point of these techniques is a nuclear energy density that is built out 
of local or non-local one-body densities~\cite{RevModPhys.75.121,schunck_energy_2019}. 
EDF techniques fall into two broad categories: 
On the one hand, within the mean-field or single-reference (SR) EDF method,
the energy kernel is calculated with densities chosen to be constructed from
a single (symmetry-breaking) auxiliary product state. 
The Hartree-Fock (HF) and Hartree-Fock-Bogolyubov (HFB) methods are
two realisations of such SR approach.
On the other hand, there are methods for which the densities entering the energy 
kernel are constructed from pairs of product states out a large 
set of reference states, and which are called beyond-mean-field or 
multi-reference (MR) methods. The restoration of symmetries and the Generator 
Coordinate Method (GCM) fall within this second type of approach.

The strategy for building the EDF is guided by general symmetry principles, 
nuclear phenomenology, and also numerical considerations.
For some forms, the \textit{ansatz} is made directly at the level of an expression 
for the energy in terms of densities. Following Ref.~\cite{sadoudi_skyrme_2013}, these
will be called \textit{general functionals} in what follows, examples being 
the Fayans~\cite{FAYANS200049} and BCPM~\cite{BALDO2008390,PhysRevC.87.064305} EDFs. 
Other forms of EDFs are constructed from, or at least motivated by, an \textit{ansatz}
for an underlying effective nucleon-nucleon interaction that serves as the generator 
of the functional. If the generator is a strict many-body operator and the energy is 
calculated as its strict many-body expectation value, the resulting EDF will be called a
\textit{pseudopotential-based functional}. Most existing functionals are however 
constructed by a mixture of these two concepts; following again 
Ref.~\cite{sadoudi_skyrme_2013}, these 
will be called \textit{hybrid functionals} in what follows.

Even describing simultaneously just some of the most basic nuclear properties 
already demands a very specific form of the EDF. While the exchange term from 
a finite-range interaction or a velocity-dependent pure two-body interaction can 
generate saturation of nuclear matter through their implicit or explicit 
momentum dependence,  
parameterisations of such interactions that provide reasonable values of the saturation 
density $\rho_{\text{sat}}$ and energy per particle $E/A$ will inevitably lead to 
an effective mass of at most about $m^*_0/m \simeq 0.4$ -- which is too small for 
a realistic description of single-particle spectra -- irrespective of the form 
of the two-body interaction \cite{f_weisskopf_problem_1957,davesne_two-body_2018}.

The EDF -- or its generator -- has therefore to be extended beyond two-body terms.
The most straightforward possibility is to include a three-nucleon interaction, 
for which there is also empirical and theoretical evidence from first 
principles \cite{Kalantar-Nayestanaki_2012,RevModPhys.85.197,HEBELER20211}. 
Saturation of nuclear matter 
can then be generated from the different growth of overall attractive two-body 
and overall repulsive three-body terms with density. For reasons of numerical cost, 
there is a clear preference for contact three-body interactions. Historically, 
the first attempt was even limited to the simplest gradientless
form~\cite{PhysRevC.5.626,BEINER197529} already suggested in
Refs.~\cite{skyrme_cvii_1956,skyrme_effective_1958}. 
Using such three-body contact force together with a two-body contact Skyrme-type
interaction was quickly found to have several severe drawbacks: for 
parameterisations that are adjusted to yield realistic values of 
$\rho_{\text{sat}}$, $E/A$, and $m^*_0/m$ such as SIII~\cite{BEINER197529}, 
the compression modulus $K_\infty$ acquires an unrealistically 
large value, the pairing interaction becomes very weak or even repulsive if the 
same interaction is used in the pairing channel \cite{AKITO198119}, and, most severely,
symmetric matter has a spin instability at saturation
density~\cite{CHANG1975205,BACKMAN1975209} as indicated by the Landau parameter 
$g_0 = -1.576$ of SIII \cite{BACKMAN1975209}. The latter makes
it impossible to consistently calculate odd- and odd-odd nuclei~\cite{PASSLER1976253,STRINGARI197687}
or rotational states. All of these problems can be linked to the very rigid 
connection between the relevant channels in the trilinear terms of the EDF 
that has just one free parameter~\cite{PhysRevC.13.1664}.

To overcome the latter of these problems without introducing a significantly more 
complicated form of EDF, the gradientless zero-range three-body pseudopotential
was replaced by a linearly density-dependent gradientless two-body interaction
with a specific relative weight $x_3$ of the spin-exchange interaction 
\cite{PhysRevC.5.626} that yields a hydrid functional with the same time-even 
trilinear terms, but leads to trilinear spin (and pairing) terms with different 
isospin structure and different signs of some coupling constants. 
To construct parameter sets with more realistic values of $K_\infty$, the linear 
power of the density entering the two-body density-dependent term was additionally 
replaced by a much smaller fractional power of the density~\cite{KOHLER1976301}, 
which can be motivated as an approximation to the effective interaction of 
Brueckner-HF theory that absorbs in-medium correlation effects. Because this 
simple minimal form of density-dependent two-body term successfully met the 
phenomenological requirements asked for at that time, it has become an ingredient 
of the standard forms of the widely-used contact Skyrme~\cite{CHABANAT1998231,1998441},
and finite-range Gogny 
EDFs~\cite{decharge_hartree-fock-bogolyubov_1980,BERGER1991365,PhysRevLett.102.242501}, 
as well as the regularised finite-range EDFs of Refs.~\cite{dobaczewski_effective_2012,raimondi_nonlocal_2014,bennaceur_nonlocal_2017} 
and the finite-range M3Y 
EDF~\cite{PhysRevC.68.014316,PhysRevC.81.027301,PhysRevC.82.029903,PhysRevC.87.014336}. 
The density-dependent terms are not always used in the same spirit, though. 
For the Gogny force, the regularised EDFs, and also recent M3Y interactions, 
and a very few Skyrme parameterisations like 
SkP~\cite{dobaczewski_hartree-fock-bogolyubov_1984},
the same density-dependent effective two-body interaction is used to generate the 
particle-hole and particle-particle parts of the EDF as the expectation value
taken for the auxiliary states, producing a hybrid EDF that is close in spirit
to the pseudopotential-based functionals. For the vast majority 
of Skyrme parameterisations, however, the pairing part of the 
EDF is constructed and adjusted independently from the particle-hole part. This 
practice is motivated by the more limited number of free parameters of 
traditional Skyrme EDFs, the necessity to regularise contact pairing
interactions~\cite{PhysRevLett.88.042504},
and the limited empirical information about the 
in-medium pairing interaction. In addition, some spin terms are also often omitted, 
which produces hybrid functionals that are close in spirit to general functionals.

Driven by additional phenomenological constraints, often concerning nuclear
response  properties or neutron-star physics, more complex forms of density
dependence such as a finite-range density dependence for the Gogny
force~\cite{chappert_nouvelles_2007,PhysRevC.91.034312}, 
and either multiple density dependencies of the gradientless
terms~\cite{lesinski_isovector_2006}
and/or density-dependent gradient terms of the Skyrme
EDF~\cite{KREWALD1977166,PhysRevC.80.065804,Grams23,PhysRevC.110.065806}
have also been considered, but are in general not widely-used outside 
of the groups that proposed them. Even less attention has been given to
the exploration of higher-order gradientful 
contact three-body interactions, either in the context of
Skyrme~\cite{LIU19759,WAROQUIER1983269,ARIMA1986286,ZHENG1990342,LIU19911}
or Gogny interactions~\cite{ONISHI1978336}, none of which became widely used.
There is, however, a renewed interest in working with true many-body 
interactions that is motivated by formal and practical difficulties that arise 
in MR EDF calculations.

Mean-field, \textit{i.e.} SR EDF, calculations are not always sufficient
to reliably describe all nuclei, and some types of states and 
observables are even not describable at all by such methods, such that 
MR techniques have to be used instead. 
Over the years, however, it has been realised, that general and hybrid
functionals are mathematically ill-defined when being used to calculate 
the -- in general complex -- energy kernels that enter MR calculations. 
There are two distinct sources of problems. First, such functionals are 
contaminated with spurious contributions that can produce discontinuities or even
divergences~\cite{ANGUIANO2001467,PhysRevC.76.054315,PhysRevC.79.044318,PhysRevC.79.044319}
when plotting energy kernels as a function of some collective coordinate.
This problem is linked to self-interaction contributions to the energy that should
not exist as a consequence of the Pauli principle~\cite{PhysRevC.79.044319}.
A second problem concerns general and hybrid functionals that contain terms that become 
non-analytical functions in the complex plane, such that the MR energy kernel has 
branch cuts and becomes multi-valued~\cite{PhysRevC.76.054315,duguet_particle-number_2009}. The prevalent example are density dependencies 
with a non-integer power as used by (most) Skyrme and Gogny EDFs. 
Some workarounds around these problems have been
proposed~\cite{ANGUIANO2001467,PhysRevC.79.044318,PhysRevC.79.044319,PhysRevC.90.054303},
but cannot be expected to work in all use cases of interest~\cite{Robledo_2007,Robledo_2010}.

As of today, the only viable option to safely bypass these pathologies is 
to work with pseudo\-po\-ten\-tial-based functionals that are strictly and exactly
constructed as many-body matrix elements of a generator between the auxiliary 
states~\cite{sadoudi_skyrme_2013} without any modification, which 
automatically generates an energy kernel that is an analytical function 
in the complex plane and for which all spurious self-interactions cancel 
out. But, for the reasons sketched above, all existing functionals that 
provide a high-quality description of nuclear phenomenology are either 
a general or a hybrid parameterisation. And even worse, the early 
attempts to construct predictive pseudopotential-based functionals with 
a simple form all failed.

Some exploratory studies towards realistic extended pseudopotential-based 
functionals for MR calculations have already been undertaken. These have to 
take into account yet another issue not already mentioned, which are finite-size 
spin and/or isospin instabilities \cite{davesne_linear_2021}, i.e.\ the 
possibility of a spurious transition of homogeneous matter to \textit{inhomogeneous} 
isospin-asymmetric or spin-polarised matter at densities and length scales 
probed by finite nuclei. This issue has to be distinguished from the instability 
for a transition towards asymmetric and/or polarised \textit{homogeneous} 
matter that is signaled by Landau parameters. Indeed, when keeping the time-odd 
terms of the Skyrme EDF with their native coupling constant, as would be done 
for a hybrid or pseudopotential-based functional, many standard parametrisations 
exhibit finite-size instabilities in the spin
channels~\cite{PhysRevC.81.024316,Pototzky2010,PhysRevC.85.014326}. An example relevant 
for our purpose of building pseudopotential-based functionals is the early 
Skyrme parametrisation SIV of Ref.~\cite{BEINER197529} that includes a weaker 
three-body force than SIII, leading to a Landau parameter of
$g_0 = 0.059$~\cite{BACKMAN1975209} that is well above the stability threshold of $-1$, 
but SIV has a finite-size spin instability instead \cite{sadoudi_skyrme_2013-1}. 
Either type of nonphysical behaviour can be avoided when adjusting parameter 
sets with a constraint on the response properties of infinite matter 
\cite{davesne_linear_2021}, which is what was done for SLyMR0 and SLyMR1.

Built for proof-of-concept spuriousity-free MR calculations, in particular of 
odd-mass nuclei~\cite{PhysRevLett.113.162501},
the SLyMR0 parametrisation~\cite{sadoudi_skyrme_2013-1} adds gradientless 
three- and four-body contact forces to the standard Skyrme two-body generator. 
For the same reasons already mentioned above for SIII and SIV, it is not 
possible to reproduce entirely realistic nuclear matter properties at 
saturation within this form of pseudopotential-based functional. 
Keeping the trilinear terms 
sufficiently small, a compromise that generates reasonable pair
correlations in nuclei can be found for which the saturation point is slightly 
off the empirical value, and spin-instabilities of either kind are pushed 
to high densities that are not probed in finite nuclei. But with
$m^{*}_0/m = 0.47$ the effective mass is only moderately larger than 
the largest possible value that can be found for pure two-body forces, 
and the symmetry energy coefficient $J$ can also not be larger than about 23~MeV. 

Within this form for the EDF, significant improvement for one nuclear matter property 
is only possible at the expense of degrading others. This stimulated the 
development and implementation of a complete central three-body contact 
interaction with two gradients \cite{sadoudi_skyrme_2013} that supersedes 
the ensemble of all ad-hoc forms considered before in 
Refs.~\cite{LIU19759,WAROQUIER1983269,ARIMA1986286,ZHENG1990342,LIU19911,ONISHI1978336}.
Even with its more flexible structure that allows for values of $\rho_{\text{sat}}$, 
$E/A$, and $J$ close to the empirical ones and to improve 
on the description of finite nuclei \cite{jodon_ajustements_2014}, with 
$m^*_0/m=0.53$ the effective mass has to remain relatively low,
and pairing correlations collapse for many nuclei, in particular 
deformed ones. The four-body term from SLyMR0, although considered 
during the parameter adjustment, was finally dropped as it did not bring a
significant improvement.
Still, the resulting parameter set SLyMR1 has nonetheless led 
to remarkable results from MR calculations of spectroscopic properties 
of some heavy nuclei~\cite{bally22,bally23}.

Constructing an EDF based on a zero-range pseudopotential that is
competitive for standard observables when compared to the widely-used 
general or hybrid EDFs will require even higher-order terms, 
if it can be achieved at all. This apparent difficulty can most probably 
be attributed to the tight interrelations between the different channels 
of the EDF that arise from both two-body and many-body contact interactions.
Indeed, a finite-range two-body interaction can give independent contributions
in the four spin and isospin channels $(S,T)$ (with $S=0$ or 1 the total
spin of the interacting nucleons and $T=0$ or 1 their total isospin).
For a contact interaction, only half of these remain linearly independent.
This significantly reduces the flexibility of the two-body interaction and
creates correlations between the different $(S,T)$ channels that might be unwelcome.

As asserted above, the only possibility to make the Skyrme EDF more flexible is to 
consider higher-order terms in gradients beyond the leading order (LO) of
gradientless terms and the next-to-leading-order (NLO) of terms
with two gradients. Work in this direction is underway at the level 
of hybrid density-dependent functionals~\cite{PhysRevC.96.044330,grams26}.

For the aforementioned reasons, even greater flexibility can be achieved at 
every order if the two-body interaction is of finite-range type.
The recent development of a flexible form of a finite-range two-body EDF
generator~\cite{dobaczewski_effective_2012,raimondi_nonlocal_2014,bennaceur_nonlocal_2017} 
actually provides the context for the present study.
It can be seen as an extension of the Brink and Boeker interaction~\cite{BRINK19671}
with derivative terms up to N$^3$LO, and also at the same time as a generalised
Skyrme interaction for which the contact form factor is replaced by a 
finite-range Gaussian one, reason for which it is called a 
\textit{regularised pseudopotential}.
Its general form will be sketched in section~\ref{subsec:2b-part-regmr3}.

As explained above, when aiming at MR calculations such two-body interaction 
has to be complemented with, at least, a three-body interaction 
that provides greater flexibility than earlier attempts.
A natural choice would be a finite-range three-body interaction constructed along 
the same lines with a Gaussian form factor that is a function of the three relative 
distances between the interacting nucleons. In an EDF context, such interaction has been 
considered for schematic calculations of very light nuclei~\cite{ZHENG1990342}, but the
unfavourable scaling of its numerical cost makes its use in systematic calculations
unfeasible, especially for studies of heavy deformed nuclei.

We mention in passing that we made several attempts to adjust a parameterisation 
that combines the regularised two-body generator with the zero-range NLO three-body 
term as used for SLyMR1, always leading to intractable problems. Indeed, we observed
that the competition between the finite-range two-body interaction (usually attractive in the
pairing channel) and the zero-range three-body terms (that are usually
repulsive in the pairing channel)
leads to a collapse of the local part of the pairing field and an unrealistically 
large attractive non-local part. This problem can only be circumvented if
the strength of the three-body interaction is made relatively small compared 
to the two-body part. However, this then always yields an effective mass 
that is not significantly larger than 0.4.

The apparent impossibility to use a fully finite-range or fully zero-range three-body
interaction lead us to consider an intermediate choice, \textit{i.e.}~a three-body interaction
which, schematically, is the product of a Gaussian of the relative distance between
two coordinates and a Dirac $\delta$ function of the relative distance between two other ones.
This type of interaction will be referred as \textit{semi-regularised} or 
\textit{semi-contact} interaction.

In an exploratory
study~\cite{lacroix_semicontact_2015}, motivated by~\cite{PhysRevC.85.037303},
we have considered a non-local form
for the interaction between three nucleons at positions $\bfr_1$, $\bfr_2$ and $\bfr_3$
built from
\begin{equation}
    g(\mathbf{r}_{12})\, \delta
    (\mathbf{r}_{3} - \mathbf{R}_{12})\,,
    \label{eq:3b-sr-bennaceur-lacroix}
\end{equation}
with $g$ a Gaussian form factor and where
$\mathbf{R}_{12} = (\mathbf{r}_{1}+\mathbf{r}_{2})/2$ is the centre-of-mass of nucleons at positions $\mathbf{r}_{1}$ and $\mathbf{r}_{2}$.
We recall here that, strictly speaking, expression~(\ref{eq:3b-sr-bennaceur-lacroix}) 
cannot represent an interaction unless it is symmetrised under
all possible exchanges of the coordinates. Equation~(\ref{eq:3b-sr-bennaceur-lacroix}) is
therefore the seed from which the interaction is formed.

This interaction has several interesting features when 
being combined with a Gaussian two-body force.
The resulting EDF proves to have a flexibility sufficient to
obtain the desired empirical properties of infinite nuclear matter.
Furthermore, despite the presence of a $\delta$ function, nucleons always interact at a
finite distance. This means that pairing energy does not diverge in HFB calculations
and therefore calculations do not require the use of a cut-off. Its form, with a dependence
on the three angles between relative coordinates makes it nonetheless very challenging to
implement in codes for finite nuclei, even in spherical symmetry.

In this work, we consider an alternative form of semi-contact three-body pseudopotential.
It is inspired
by the expression~(\ref{eq:3b-sr-bennaceur-lacroix}) and is built from the symmetrised version of
\begin{equation}
    g(\bfr_{12})\, \delta(\bfr_{23})\,,   \label{eq:3b-sr}
\end{equation}
where $\bfr_{ij}=\bfr_j-\bfr_i$\,.
This interaction has a spatial structure
simpler than~(\ref{eq:3b-sr-bennaceur-lacroix}) since two nucleons out of the three interact when
they are at the same position. This can lead to an EDF which depends on the local
anomalous density and therefore would require the use of a cut-off to prevent the 
divergence of
the energy~\cite{PhysRevLett.88.042504}. 
However, as will be shown, a special condition on its spin structure can
be used to avoid this divergence without using a cut-off.

This article is organised as follows. In
section~\ref{sec:pn-representation}, we briefly recall
the definition of the one-body densities when proton-neutron mixing is allowed or not.
In section~\ref{subsec:2b-part-regmr3}, we 
recall the form of the two-body regularised local interaction already introduced
in~\cite{dobaczewski_effective_2012}.
Section~\ref{sec:construction-pseudo-potential} dives 
into the process that 
allows to construct the most general form of the three-body
semi-contact force which is usable in HFB calculations without the introduction of a cut-off.
In section~\ref{sec:regmr3-parametrization}, we will
summarise results obtained with the RegMR3 functional that explored a
special case of the three-body semi-contact force~\cite{costa_interactions_2022},
and discuss its limitations that motivated the extended form presented here.
Finally, the mathematical expression of the EDF will be provided. 
Sections~\ref{sec:relations-INM} and~\ref{sec:spherical-symmetry} are
intended to give necessary relations used in practical implementation.
They respectively discuss homogeneous infinite nuclear matter and calculations
for spherically-symmetric systems.


%
\section{Proton-neutron representation\label{sec:pn-representation}}

The SR EDF obtained from the three-body interaction will be derived considering
that all possible symmetries of the nuclear Hamiltonian operator might be broken
by the auxiliary states.
In particular,
for completeness and to facilitate future developments,
we will not assume that the single particle-states are 
eigenstates of the isospin Pauli matrix $\tau_{3}$.
Following the notations used in~\cite{perlinska_local_2004}, we use here
the ``breve'' representation for the
pair density matrices that are needed to build the most general isoscalar 
and isovector pair energy density of proton-neutron-mixed quasiparticle vacua,
and later give the relations in the more standard ``tilde'' representation 
of neutron and proton pair density matrices that is more 
straightforward to use when proton-neutron mixing is neglected. The particle-hole
density matrices can be expanded on identity and spin and isospin Pauli matrices
using the scalar-isoscalar $\rho_0$, scalar-isovector $\vec\rho$, vector-isoscalar $\mathbf s_0$,
and vector-isovector $\vec{\mathbf s}$ densities as follows
\begin{align}
    \rho (\mathbf{r}_{1} s_{1} q_{1}, \mathbf{r}_{2} s_{2} q_{2})
    &= \braket{\Phi | a^{\dagger}_{\mathbf{r}_{2} s_{2} q_{2}}
    a_{\mathbf{r}_{1} s_{1} q_{1}} | \Phi}\,, \nonumber \\
    & = \frac{1}{4} \big[
    \rho_{0} (\mathbf{r}_{1}, \mathbf{r}_{2}) \delta_{s_{1}s_{2}}
    \delta_{q_{1}q_{2}}
    + \delta_{s_{1}s_{2}} \vec{\rho} (\mathbf{r}_{1}, \mathbf{r}_{2})
    \circ \vec{\tau}_{q_{1} q_{2}} \nonumber \\
    &\hskip 3.5cm+ \mathbf{s}_{0} (\mathbf{r}_{1}, \mathbf{r}_{2}) \cdot
    \bm{\sigma}_{s_{1}s_{2}} \delta_{q_{1}q_{2}}
    + \vec{\mathbf{s}} (\mathbf{r}_{1}, \mathbf{r}_{2}) \cdot 
    \bm{\sigma}_{s_{1}s_{2}} \circ \vec{\tau}_{q_{1}q_{2}}
    \big]\,.
    \label{eq:decomposition-normal-density}
\end{align}
The particle-particle density can be expanded in the same way as
\begin{align}
    \breve{\rho} (\mathbf{r}_{1} s_{1} q_{1}, \mathbf{r}_{2} s_{2} q_{2})
    &= 4 s_{2} q_{2} \braket{\Phi | a_{\mathbf{r}_{2} -s_{2} -q_{2}}
    a_{\mathbf{r}_{1} s_{1} q_{1}} | \Phi}\,,
    \nonumber \\
    & = \frac{1}{4} \big[
    \breve{\rho}_{0} (\mathbf{r}_{1}, \mathbf{r}_{2}) \delta_{s_{1}s_{2}}
    \delta_{q_{1}q_{2}}
    + \delta_{s_{1}s_{2}} \vec{\breve{\rho}} (\mathbf{r}_{1}, \mathbf{r}_{2})
    \circ \vec{\tau}_{q_{1} q_{2}} \nonumber \\
    &\hskip 3.5cm
    + \breve{\mathbf{s}}_{0} (\mathbf{r}_{1}, \mathbf{r}_{2}) \cdot
    \bm{\sigma}_{s_{1}s_{2}} \delta_{q_{1}q_{2}}
    + \vec{\breve{\mathbf{s}}} (\mathbf{r}_{1}, \mathbf{r}_{2}) \cdot 
    \bm{\sigma}_{s_{1}s_{2}} \circ \vec{\tau}_{q_{1}q_{2}}
    \big]\,,
    \label{eq:decomposition-breve-density}
\end{align}
where $\ket{\Phi}$ is an independent quasiparticle state. Bold symbols and
arrows indicate vectors in position and isospin space, respectively. 
The corresponding scalar products are respectively noted with the symbols ``$\cdot$''
and ``$\circ$''
(see also appendix~\ref{app:functional-with-pn} for more details about  notations).
The densities in the above equations are defined as 
\begin{align}
    \rho_{0} (\mathbf{r}_{1}, \mathbf{r}_{2})
    & = \sum_{sq} \rho (\mathbf{r}_{1} s q, \mathbf{r}_{2} s q)\,, 
    \label{eq:definition-scalar-isoscalar-density-with-pn-mixing} \\
    \vec{\rho} (\mathbf{r}_{1}, \mathbf{r}_{2})
    & = \sum_{sq_{1}q_{2}}
    \rho (\mathbf{r}_{1} s q_{1}, \mathbf{r}_{2} s q_{2})
    \vec{\tau}_{q_{2}q_{1}}\,,
    \label{eq:definition-scalar-isovector-density-with-pn-mixing} \\
    \mathbf{s}_{0} (\mathbf{r}_{1}, \mathbf{r}_{2})
    & = \sum_{s_{1}s_{2}q}
    \rho (\mathbf{r}_{1} s_{1} q, \mathbf{r}_{2} s_{2} q)
    \bm{\sigma}_{s_{2}s_{1}}\,,
    \label{eq:definition-vector-isoscalar-density-with-pn-mixing} \\
    \vec{\mathbf{s}} (\mathbf{r}_{1}, \mathbf{r}_{2})
    & = \sum_{s_{1}s_{2}q_{1}q_{2}}
    \rho (\mathbf{r}_{1} s_{1} q_{1}, \mathbf{r}_{2} s_{2} q_{2})
    \bm{\sigma}_{s_{2}s_{1}} \vec{\tau}_{q_{2}q_{1}}\,,
    \label{eq:definition-vector-isovector-density-with-pn-mixing}
\end{align}
and 
\begin{align}
    \breve{\rho}_{0} (\mathbf{r}_{1}, \mathbf{r}_{2})
    & = \sum_{sq} \breve{\rho} (\mathbf{r}_{1} s q, \mathbf{r}_{2} s q)\,, 
    \label{eq:definition-scalar-isoscalar-breve-density} \\
    \vec{\breve{\rho}} (\mathbf{r}_{1}, \mathbf{r}_{2})
    & = \sum_{sq_{1}q_{2}}
    \breve{\rho} (\mathbf{r}_{1} s q_{1}, \mathbf{r}_{2} s q_{2})
    \vec{\tau}_{q_{2}q_{1}}\,,
    \label{eq:definition-scalar-isovector-breve-density} \\
    \breve{\mathbf{s}}_{0} (\mathbf{r}_{1}, \mathbf{r}_{2})
    & = \sum_{s_{1}s_{2}q}
    \breve{\rho} (\mathbf{r}_{1} s_{1} q, \mathbf{r}_{2} s_{2} q)
    \bm{\sigma}_{s_{2}s_{1}}\,,
    \label{eq:definition-vector-isoscalar-breve-density} \\
    \vec{\breve{\mathbf{s}}} (\mathbf{r}_{1}, \mathbf{r}_{2})
    & = \sum_{s_{1}s_{2}q_{1}q_{2}}
    \breve{\rho} (\mathbf{r}_{1} s_{1} q_{1}, \mathbf{r}_{2} s_{2} q_{2})
    \bm{\sigma}_{s_{2}s_{1}} \vec{\tau}_{q_{2}q_{1}}\,.
    \label{eq:definition-vector-isovector-breve-density}
\end{align}
%
%
When proton-neutron mixing is not considered,
only the zeroth and third
isospin components of the density~\eqref{eq:decomposition-normal-density}
and the first and second isospin components of the
density~\eqref{eq:decomposition-breve-density} are
non-zero~\cite{perlinska_local_2004}.
It is then easier to work with the pair density expressed in the ``tilde''
representation, namely 
\begin{equation}
    \tilde{\rho} (\mathbf{r}_{1} s_{1} q_{1}, \mathbf{r}_{2} s_{2} q_{2})
    = 2 q_{2} \breve{\rho}
    (\mathbf{r}_{1} s_{1} q_{1}, \mathbf{r}_{2} s_{2} -\!q_{2})\,.
    \label{eq:link-rho-tilde-and-rho-breve}
\end{equation}
In that case, the decomposition~\eqref{eq:decomposition-normal-density}
becomes
\begin{align}
    \rho (\mathbf{r}_{1}s_{1}q_{1}, \mathbf{r}_{2}s_{2}q_{2})
    & = \frac{1}{4} \big[
    \rho_{0} (\mathbf{r}_{1}, \mathbf{r}_{2})
    \delta_{s_{1}s_{2}} \delta_{q_{1}q_{2}}
    + \rho_{1} (\mathbf{r}_{1}, \mathbf{r}_{2})
    \delta_{s_{1}s_{2}} (\tau_{3})_{q_{1}q_{2}}
    \nonumber \\
    & \hspace{0.4cm}
    + \mathbf{s}_{0} (\mathbf{r}_{1}, \mathbf{r}_{2})
    \cdot \bm{\sigma}_{s_{1}s_{2}} \delta_{q_{1}q_{2}}
    + \mathbf{s}_{1} (\mathbf{r}_{1}, \mathbf{r}_{2})
    \cdot \bm{\sigma}_{s_{1}s_{2}}
    (\tau_{3})_{q_{1}q_{2}}
    \big]\,,
    \label{eq:decomposition-one-body-density-matrix-without-pn-mixing}
\end{align}
(with, as is customary, the third components
of $\vec\rho$ and $\vec{\mathbf s}$ denoted $\rho_1$ and $\mathbf s_1$)
and, the anomalous densities for each species label with
$q=n$ or $p$
\begin{align}
    \tilde{\rho}_{q} (\mathbf{r}_{1} s_{1}, \mathbf{r}_{2} s_{2})
    = \frac{1}{2} \big[
    \tilde{\rho}_{q} (\mathbf{r}_{1}, \mathbf{r}_{2}) \delta_{s_{1}s_{2}}
    + \tilde{\mathbf{s}}_{q} (\mathbf{r}_{1}, \mathbf{r}_{2}) \cdot 
    \bm{\sigma}_{s_{1}s_{2}}
    \big]\,,
    \label{eq:decomposition-one-body-tilde-density-matrix}
\end{align}
is used instead of~\eqref{eq:decomposition-breve-density}. The
definitions~\eqref{eq:definition-scalar-isoscalar-density-with-pn-mixing}
and~\eqref{eq:definition-vector-isoscalar-density-with-pn-mixing} remain 
unchanged while
expressions~\eqref{eq:definition-scalar-isovector-density-with-pn-mixing}
and~\eqref{eq:definition-vector-isovector-density-with-pn-mixing} reduce to
\begin{align}
    \rho_{1} (\mathbf{r}_{1}, \mathbf{r}_{2})
    & = \sum_{sq_{1}q_{2}}
    \rho (\mathbf{r}_{1} s q_{1}, \mathbf{r}_{2} s q_{2})
    (\tau_{3})_{q_{2}q_{1}}\,,
    \label{eq:definition-scalar-isovector-density-without-pn-mixing} \\
    \mathbf{s}_{1} (\mathbf{r}_{1}, \mathbf{r}_{2})
    & = \sum_{s_{1}s_{2}q_{1}q_{2}}
    \rho (\mathbf{r}_{1} s_{1} q_{1}, \mathbf{r}_{2} s_{2} q_{2})
    \bm{\sigma}_{s_{2}s_{1}} (\tau_{3})_{q_{2}q_{1}}\,.
    \label{eq:definition-vector-isovector-density-without-pn-mixing}
\end{align}
Note that in
equations~(\ref{eq:decomposition-one-body-density-matrix-without-pn-mixing}),
(\ref{eq:definition-scalar-isovector-density-without-pn-mixing}),
and~(\ref{eq:definition-vector-isovector-density-without-pn-mixing}),
as already mentioned, we note $\rho_{1}$ instead of $\rho_{3}$ and 
$\mathbf{s}_{1}$ instead of $\mathbf{s}_{3}$.

The non-vanishing components of the
densities~\eqref{eq:definition-scalar-isovector-breve-density}
and~\eqref{eq:definition-vector-isovector-breve-density} give the scalar
and vector parts of the density from~\eqref{eq:decomposition-one-body-tilde-density-matrix}
\begin{align}
    \tilde{\rho}_{q} (\mathbf{r}_{1}, \mathbf{r}_{2})
    & = \sum_{s} \tilde{\rho}_{q} (\mathbf{r}_{1} s, \mathbf{r}_{2} s)\,, 
    \label{eq:definition-density-rho-tilde-q} \\
    \tilde{\mathbf{s}}_{q} (\mathbf{r}_{1}, \mathbf{r}_{2})
    & = \sum_{s_{1}s_{2}}
    \tilde{\rho}_{q} (\mathbf{r}_{1} s_{1}, \mathbf{r}_{2} s_{2})
    \bm{\sigma}_{s_{2}s_{1}}\,.
    \label{eq:definition-density-s-tilde-q}
\end{align}
Using relations~\eqref{eq:definition-scalar-isoscalar-density-with-pn-mixing}-\eqref{eq:definition-vector-isovector-breve-density}
and~\eqref{eq:definition-scalar-isovector-density-without-pn-mixing}-\eqref{eq:definition-density-s-tilde-q}
combined with the definition~\eqref{eq:link-rho-tilde-and-rho-breve}
allows to make the link between both representations when proton-neutron
mixing is not considered, as reported on 
table~\ref{tab:densities-from-tilde-to-breve-representation}.

%
%

\begin{table}[htbp]
    \centering
    \caption{\label{tab:densities-from-tilde-to-breve-representation}
    Relations between the ``tilde'' and ``breve'' representations when
    proton-neutron mixing is not considered.}
    \begin{tabular}{lcl}
        \toprule
        \multicolumn{1}{c}{
        Normal densities} & ~ & \multicolumn{1}{c}{Pairing densities} \\
        \midrule 
        $\rho_{0} (\mathbf{r}_{1}, \mathbf{r}_{2})
        = \rho_{\text{n}} (\mathbf{r}_{1}, \mathbf{r}_{2})
        + \rho_{\text{p}} (\mathbf{r}_{1}, \mathbf{r}_{2})$
        & & $\breve{\rho}_{0} (\mathbf{r}_{1}, \mathbf{r}_{2}) = 0$ \\
        $\rho_{1} (\mathbf{r}_{1}, \mathbf{r}_{2}) = 0$ 
        & & $\breve{\rho}_{1} (\mathbf{r}_{1}, \mathbf{r}_{2})
        = \tilde{\rho}_{\text{n}} (\mathbf{r}_{1}, \mathbf{r}_{2})
        - \tilde{\rho}_{\text{p}} (\mathbf{r}_{1}, \mathbf{r}_{2})$ \\
        $\rho_{2} (\mathbf{r}_{1}, \mathbf{r}_{2}) = 0$
        & & $\breve{\rho}_{2} (\mathbf{r}_{1}, \mathbf{r}_{2})
        = \mathrm{i}\, \big[
        \tilde{\rho}_{\text{n}} (\mathbf{r}_{1}, \mathbf{r}_{2})
        + \tilde{\rho}_{\text{p}} (\mathbf{r}_{1}, \mathbf{r}_{2})
        \big]$ \\
        $\rho_{3} (\mathbf{r}_{1}, \mathbf{r}_{2})
        = \rho_{\text{n}} (\mathbf{r}_{1}, \mathbf{r}_{2})
        - \rho_{\text{p}} (\mathbf{r}_{1}, \mathbf{r}_{2})$
        & & $\breve{\rho}_{3} (\mathbf{r}_{1}, \mathbf{r}_{2}) = 0$ \\
        \midrule
        %
        $\mathbf{s}_{0} (\mathbf{r}_{1}, \mathbf{r}_{2})
        = \mathbf{s}_{\text{n}} (\mathbf{r}_{1}, \mathbf{r}_{2})
        + \mathbf{s}_{\text{p}} (\mathbf{r}_{1}, \mathbf{r}_{2})$
        & & $\breve{\mathbf{s}}_{0} (\mathbf{r}_{1}, \mathbf{r}_{2})
        = \mathbf{0}$ \\
        $\mathbf{s}_{1} (\mathbf{r}_{1}, \mathbf{r}_{2})
        = \mathbf{0}$ 
        & & $\breve{\mathbf{s}}_{1} (\mathbf{r}_{1}, \mathbf{r}_{2})
        = \tilde{\mathbf{s}}_{\text{n}} (\mathbf{r}_{1}, \mathbf{r}_{2})
        - \tilde{\mathbf{s}}_{\text{p}} (\mathbf{r}_{1}, \mathbf{r}_{2})$ \\
        $\mathbf{s}_{2} (\mathbf{r}_{1}, \mathbf{r}_{2})
        = \mathbf{0}$
        & & $\breve{\mathbf{s}}_{2} (\mathbf{r}_{1}, \mathbf{r}_{2})
        = \mathrm{i}\, \big[
        \tilde{\mathbf{s}}_{\text{n}} (\mathbf{r}_{1}, \mathbf{r}_{2})
        + \tilde{\mathbf{s}}_{\text{p}} (\mathbf{r}_{1}, \mathbf{r}_{2})
        \big]$ \\
        $\mathbf{s}_{3} (\mathbf{r}_{1}, \mathbf{r}_{2})
        = \mathbf{s}_{\text{n}} (\mathbf{r}_{1}, \mathbf{r}_{2})
        - \mathbf{s}_{\text{p}} (\mathbf{r}_{1}, \mathbf{r}_{2})$
        & & $\breve{\mathbf{s}}_{3} (\mathbf{r}_{1}, \mathbf{r}_{2})
        = \mathbf{0}$ \\
        \bottomrule
    \end{tabular}
\end{table}



\section{Construction of the pseudopotential\label{sec:construction-pseudo-potential}}

\subsection{Two-body part of the pseudopotential\label{subsec:2b-part-regmr3}}

The motivation for the introduction of a two-body regularised potential
in its most general form and the derivation of the corresponding contributions to the
EDF has been discussed in previous
works~\cite{dobaczewski_effective_2012,raimondi_nonlocal_2014,bennaceur_nonlocal_2017,bennaceur_new_2014}
but is partly repeated here for completeness.

The most general form of the central part of a finite-range
two-body $\text{N}^{p}\text{LO}$ pseudopotential, \textit{i.e.}~a pseudopotential
regularised at order $p \in \mathbb{N}$, is a linear combination of terms of the form 
\begin{align}
    \hat{V}_{2,j,bc}^{(n)} (\bfr_1, \bfr_2 ; \bfr_3, \bfr_4)
    = W^{(n)}_{2,j,bc}
    \hat{P}^{\sigma}_{b} \hat{P}^{\tau}_{c}
    \hat{O}_{j}^{(n)} (\mathbf{k}_{12}, \mathbf{k}_{34})
    \delta (\mathbf{r}_{13}) \delta (\mathbf{r}_{24})
    G_{a} (\mathbf{r}_{12})\,,
    \label{eq:general-form-non-local-fr-2b-NpLO-pseudo-potential}
\end{align}
with $n=2p$ and where $\mathbf{r}_{i_1i_2} = \mathbf{r}_{i_2} - \mathbf{r}_{i_1}$ are
relative positions between nucleons and $\mathbf{k}_{i_1i_2}
= \frac{1}{2 \mathrm{i}} (\bm{\nabla}_{i_2} - \bm{\nabla}_{i_1})$ their 
relative momenta. The parameters $W^{(n)}_{2,j,bc} \in \mathbb{R}$
are the strengths of the different terms of the potential indexed by $j$.
The form factor $G_{a}$ characterises the range of the 
interaction through a real parameter $a>0$ with the condition that it
reduces to a Dirac $\delta$ function for $a\to0$.
The index
$j$ lists the number of hermitian scalar operators $\hat{O}^{(n)}_{j}$ of order $n$
that can be formed from the relative momenta $\mathbf{k}_{12}^{*}$ and $\mathbf{k}_{34}$. 
Finally, $\hat{P}^{\sigma}_{b}$ and $\hat{P}^{\tau}_{c}$ are operators acting on the spin and isospin coordinates.
They are indexed by the subscripts $b,\,c \in \{ 1, 2 \}$ and respectively belong to the
sets ${\cal I}^{\sigma}_2
= \{ \hat{P}^{\sigma}_{1}=\mathbb{1}^{\sigma},\,
\hat{P}^{\sigma}_{2}=\hat{P}^{\sigma}_{12} \}$
and ${\cal I}^{\tau}_2
= \{ \hat{P}^{\tau}_{1} = \mathbb{1}^{\tau},\,
\hat{P}^{\tau}_{2}= -\hat{P}^{\tau}_{12} \}$~\cite{dobaczewski_effective_2012}.
Note that the operator $\hat{P}^{\sigma}_{12}$ acts on a ket by exchanging
the spin projection of states at position 1 and 2 but not necessarily
the spin projection of the states indexed by 1 and 2, {\it i.e.}
\begin{equation}
\hat{P}^{\sigma}_{12}|\bfr_as_aq_a,\bfr_bs_bq_b\rangle
=|\bfr_as_bq_a,\bfr_bs_aq_b\rangle\,.
\end{equation}
An equivalent definition is used for the operator $\hat{P}^{\tau}_{2}$\,.

After summation over $b$ and $c$, the above potential is usually
written~\cite{bennaceur_nonlocal_2017}
\begin{align}
     \hat{V}_{2,j}^{(n)} (\bfr_1, \bfr_2 ; \bfr_3, \bfr_4)
    &= \sum_{\hat{P}^{\sigma}_{b} \in {\cal I}^{\sigma}_2}
    \sum_{\hat{P}^{\tau}_{c} \in {\cal I}^{\tau}_2}
    \hat{V}_{2,j,bc}^{(n)} (\bfr_1, \bfr_2 ; \bfr_3, \bfr_4)\,,
    \nonumber \\
    & = \left[
    W_{j}^{(n)} \hat{\mathbb{1}}^{\sigma} \hat{\mathbb{1}}^{\tau}
    + B_{j}^{(n)} \hat{P}^{\sigma}_{12} \hat{\mathbb{1}}^{\tau}
    - H_{j}^{(n)} \hat{\mathbb{1}}^{\sigma} \hat{P}^{\tau}_{12}
    - M_{j}^{(n)}
    \hat{P}^{\sigma}_{12} \hat{P}^{\tau}_{12}
    \right] \nonumber \\
    & \hskip 1cm\times
    \hat{O}_{j}^{(n)} (\mathbf{k}_{12}, \mathbf{k}_{34})
    \delta (\mathbf{r}_{13}) \delta (\mathbf{r}_{24}) 
    G_{a} (\mathbf{r}_{12})\,,
    \label{eq:general-form-non-local-fr-2b-NpLO-pseudo-potential-WBHM}
\end{align}
with the notations for the strengths $W^{(n)}_{2,j,11} = W^{(n)}_{j}$,
$W^{(n)}_{2,j,21} = B^{(n)}_{j}$,
$W^{(n)}_{2,j,12} = H^{(n)}_{j}$,
and $W^{(n)}_{2,j,22} = M^{(n)}_{j}$.

At order $p=0$ of the potential, there exists only one scalar operator
$\hat{O}_{1}^{(0)} = \hat{\mathbb{1}}$.
The potential can then be written without the use of the gradient operators
and reads~\cite{bennaceur_nonlocal_2017}
\begin{equation}
    \hat{V}_{2,1}^{(0)} (\bfr_1, \bfr_2 ; \bfr_3, \bfr_4)
    = \left[
    W_{1}^{(0)} \hat{\mathbb{1}}^{\sigma} \hat{\mathbb{1}}^{\tau}
    + B_{1}^{(0)} \hat{P}^{\sigma}_{12} \hat{\mathbb{1}}^{\tau}
    - H_{1}^{(0)} \hat{\mathbb{1}}^{\sigma} \hat{P}^{\tau}_{12}
    - M_{1}^{(0)} \hat{P}^{\sigma}_{12} \hat{P}^{\tau}_{12}
    \right] \delta (\mathbf{r}_{13}) \delta (\mathbf{r}_{24}) 
    G_{a} (\mathbf{r}_{12})\,.
    \label{eq:general-form-local-fr-2b-LO-pseudo-potential}
\end{equation}
For orders $p>0$, the interaction can be restricted to the case of a pseudopotential which is local
as it was the case in a previous work~\cite{bennaceur_nonlocal_2017}.
The potential is local when the derivative operators commute with the Dirac $\delta$ and
hence act directly on the form factor $G_{a}$ yielding
\begin{align}
    \hat{V}_{\text{loc}}^{(n)} (\bfr_1, \bfr_2 ; \bfr_3, \bfr_4)
    &= \left[
    W_{1}^{(n)} \hat{\mathbb{1}}^{\sigma} \hat{\mathbb{1}}^{\tau}
    + B_{1}^{(n)} \hat{P}^{\sigma}_{12} \hat{\mathbb{1}}^{\tau}
    - H_{1}^{(n)} \hat{\mathbb{1}}^{\sigma} \hat{P}^{\tau}_{12}
    - M_{1}^{(n)} \hat{P}^{\sigma}_{12} \hat{P}^{\tau}_{12}
    \right] \nonumber \\
    &\hskip 1cm\times\delta (\mathbf{r}_{13}) \delta (\mathbf{r}_{24}) 
    \left( \frac{1}{2} \right)^{\frac{n}{2}} \mathbf{k}^{n}_{12}
    G_{a} (\mathbf{r}_{12})\,.
    \label{eq:general-form-local-fr-2b-NpLO-pseudo-potential}
\end{align}
If $G_{a}$ is chosen to be a normalised Gaussian form factor
\begin{equation}
    G_{a} (\mathbf{r}_{ij})
    \equiv g_{a} (\mathbf{r}_{ij})
    = \frac{\mathrm{e}^{-\frac{\mathbf{r}_{ij}^{2}}{a^{2}}}}{
    (a \sqrt{\pi})^{3}}\,,
    \label{eq:normalized-gaussian-ff}
\end{equation}
one has the property~\cite{bennaceur_nonlocal_2017}
\begin{equation}
    \left( \frac{1}{2} \right)^{\frac{n}{2}} \mathbf{k}^{n}_{ij}
    g_{a} (\mathbf{r}_{ij})
    = \left( - \frac{1}{a} \frac{\partial}{\partial a} \right)^{p}
    g_{a} (\mathbf{r}_{ij})\,,
    \label{eq:propriete-derivee-facteur-de-forme-gaussien}
\end{equation}
that substantially simplifies calculations since results at
$\text{N}^{p}\text{LO}$ are then obtained by successive application of the 
operator $-\frac{1}{a} \partial_{a}$ on LO expressions.

\subsection{Three-body part of the pseudopotential\label{subsec:3b-part-regmr3}}

A general expression for the general central part of a finite-range 
three-body pseudopotential can be written in a similar fashion as the 
form~\eqref{eq:general-form-non-local-fr-2b-NpLO-pseudo-potential} given 
for a two-body interaction. This expression will be given below in this
paragraph. For now, we only point out
that the form factor $G_{a}$ must be modified in order to include the
relative positions of the three interacting nucleons\footnote{Proper
symmetrisation of the potential also becomes relevant as it will be discussed
later.}, therefore
$G_{a}$ is changed to  $\mathcal{G}_{a}
= \mathcal{G}_{a} (\mathbf{r}_{12}, \mathbf{r}_{13}, \mathbf{r}_{23})$. 
As already mentioned in the introduction, if
$\mathcal{G}_{a}$ has finite-range in every direction, the integration
volume entering the EDF will be $\mathbb{R}^{9}$, making numerical calculations
very time consuming and possibly even intractable
for deformed nuclei. For this reason, 
we chose to discard the use of a fully finite-range regulator.

In order to reduce the dimension of the integration volume, an alternative
simpler form of the function $\mathcal{G}_{a}$ can be considered, namely 
\begin{equation}
    \mathcal{G}_{a} (\mathbf{r}_{12}, \mathbf{r}_{13}, \mathbf{r}_{23})
    = g_{a} (\mathbf{r}_{12}) \delta (\mathbf{r}_{23})\,,
    \label{eq:spatial-part-regmr3}
\end{equation}
where $g_{a}$ is the normalised Gaussian form
factor~\eqref{eq:normalized-gaussian-ff}.
Additionally, if no derivative operators are considered, the 
resulting pseudopotential is a local LO interaction. 
Because it contains both a finite- and a zero-range part, this potential 
is referred to as ``semi-contact'' or
``semi-regularised''~\cite{lacroix_semicontact_2015,costa_interactions_2022}. 
Thanks to the presence of the Dirac $\delta$ function, the dimensionality of
the problem is reduced to the evaluation of integrals in $\mathbb{R}^{6}$.
Indeed, the associated contributions to the EDF will contain, at most, products of two 
non-local densities and a local one.
This structure is thus similar to that obtained with the finite-range
density-dependent term of the D2
parametrisation of the Gogny interaction~\cite{chappert_nouvelles_2007,PhysRevC.91.034312}.
Because this latter parametrisation has already been implemented and used
in axially-symmetric HFB solvers, we expect that
it is possible to do the same with a three-body force whose spatial generator is
given by~\eqref{eq:spatial-part-regmr3}.
Furthermore, the implementation of this interaction for a
spherically-symmetric system is drastically facilitated because only the
angle between two coordinates appears (compared to the
three angles that would be involved with a complete finite-range
parametrisation of the function $\mathcal{G}_{a}$).

%
%

The generalisation of
eq.~\eqref{eq:general-form-non-local-fr-2b-NpLO-pseudo-potential} for a 
three-body potential reads
\begin{align}
    \hat{V}_{3,j,bc}^{(n)} (\bfr_1, \bfr_2, \bfr_3; \bfr_4, \bfr_5, \bfr_6)
    &= W^{(n)}_{3,j,bc}
    \hat{\cal P}^{\sigma}_{b}
    \hat{\cal P}^{\tau}_{c}
    \hat{\mathcal{O}}_{j}^{(n)}
    (\mathbf{k}_{12}, \mathbf{k}_{13}, \mathbf{k}_{23};
    \mathbf{k}_{45}, \mathbf{k}_{46}, \mathbf{k}_{56}) \nonumber \\
    &\hskip 1cm\times
    \delta (\mathbf{r}_{14}) \delta (\mathbf{r}_{25})
    \delta (\mathbf{r}_{36})
    \mathcal{G}_{a} (\mathbf{r}_{12}, \mathbf{r}_{13}, \mathbf{r}_{23})\,,
    \label{eq:general-form-non-local-fr-3b-NpLO-pseudo-potential}
\end{align}
where, again, $j$ labels the scalar operators
$\hat{\mathcal{O}}^{(n)}_{j}$ that can be formed from the relative
momenta of the nucleons. The indices $b$, $c$ allow
to keep track of the possibilities of exchanging the spin and isospin of the
three particles. There are six possible permutations for the spin projections
of three nucleons, which correspond to the operators
$\hat{\mathbb{1}}^{\sigma}$\,, $\hat{P}^{\sigma}_{23}$\,,
$\hat{P}^{\sigma}_{23}\hat{P}^{\sigma}_{12}$\,,
$\hat{P}^{\sigma}_{12}$\,,
$\hat{P}^{\sigma}_{12} \hat{P}^{\sigma}_{23}$\,,
and $\hat{P}^{\sigma}_{13}$\,. It is easy to check that the operators
$\hat{P}^{\sigma}_{23}\hat{P}^{\sigma}_{12}$ and
$\hat{P}^{\sigma}_{12} \hat{P}^{\sigma}_{23}$ are not self-adjoint but Hermitian conjugate
to one another.
Therefore, since the interaction must be self-adjoint, the spin operators
$\hat{\cal P}^\sigma_b$ from eq.~(\ref{eq:general-form-non-local-fr-3b-NpLO-pseudo-potential})
belong to the set
\begin{align}
    {\cal I}^{\sigma}_3= \left\{
    \hat{\mathcal{P}}^{\sigma}_{1} = \hat{\mathbb{1}}^{\sigma}\,,~
    \hat{\mathcal{P}}^{\sigma}_{2} = \hat{P}^{\sigma}_{12}\,,~
    \hat{\mathcal{P}}^{\sigma}_{3} = \hat{P}^{\sigma}_{13}\,,~
    \hat{\mathcal{P}}^{\sigma}_{4} = \hat{P}^{\sigma}_{23}\,,~
    \hat{\mathcal{P}}^{\sigma}_{5}
    =\tfrac{1}{2}\left(\hat{P}^{\sigma}_{23} \hat{P}^{\sigma}_{12}
    +\hat{P}^{\sigma}_{12} \hat{P}^{\sigma}_{23}\right)~
    \right\}\,.
    \label{eq:ordered-set-I-sigma-3b}
\end{align}
For a similar reason, the isospin operators $\hat{\cal P}^\tau_c$ belong to the set
\begin{align}
    {\cal I}^{\tau}_3= \left\{
    \hat{\mathcal{P}}^{\tau}_{1} = \hat{\mathbb{1}}^{\tau}\,,~
    \hat{\mathcal{P}}^{\tau}_{2} = \hat{P}^{\tau}_{12}\,,~
    \hat{\mathcal{P}}^{\tau}_{3} = \hat{P}^{\tau}_{13}\,,~
    \hat{\mathcal{P}}^{\tau}_{4} = \hat{P}^{\tau}_{23}\,,~
    \hat{\mathcal{P}}^{\tau}_{5}
    =\tfrac{1}{2}\left(\hat{P}^{\tau}_{23} \hat{P}^{\tau}_{12}
    +\hat{P}^{\tau}_{12} \hat{P}^{\tau}_{23}\right)~
    \right\}\,.
    \label{eq:ordered-set-I-tau-3b}
\end{align}
In addition, since we limit this study to a leading-order potential ($n=0$), there is only one operator
$\hat{\mathcal{O}}^{(0)}_{j}$ which is
$\hat{\mathcal{O}}^{(0)}_{1}
= \hat{\mathbb{1}}$\,. Furthermore, if one 
considers a semi-regularised interaction for which $\mathcal{G}_{a}$ is 
given by~\eqref{eq:spatial-part-regmr3}, then 
eq.~\eqref{eq:general-form-non-local-fr-3b-NpLO-pseudo-potential} reads 
\begin{align}
    \hat{V}_{3,bc} (\bfr_1, \bfr_2, \bfr_3; \bfr_4, \bfr_5, \bfr_6)
    = W_{3,bc}
    \delta (\mathbf{r}_{14}) \delta (\mathbf{r}_{25})
    \delta (\mathbf{r}_{36})
    \hat{\mathcal{P}}^{\sigma}_{b}
    \hat{\mathcal{P}}^{\tau}_{c}
    g_{a} (\mathbf{r}_{12}) \delta (\mathbf{r}_{23})\,,
    \label{eq:general-form-local-fr-3b-LO-pseudo-potential}
\end{align}
The general form of the pseudopotential can be built from
the 25 Hermitian combinations of spin and
isospin operators $\hat{\mathcal{P}}^{\sigma}_{b}
    \hat{\mathcal{P}}^{\tau}_{c}\in {\cal I}_3^{\sigma}\otimes {\cal I}_3^{\tau}$. 

Before diving into the process that further reduces the size of
${\cal I}_3^{\sigma}\otimes {\cal I}_3^{\tau}$
and allows to identify linearly independent terms to build the functional,
we emphasise that the
interaction~\eqref{eq:general-form-local-fr-3b-LO-pseudo-potential} must be
symmetrised due to the indistinguishable nature of the nucleons.
This can be achieved in two ways: (i) either symmetrisation is done by
summation of potentials obtained by performing all possible
permutations of indices 1, 2, 3
(with the same permutations on indices 4, 5, 6) in eq.~\eqref{eq:general-form-local-fr-3b-LO-pseudo-potential}
or (ii) indices $1,\,\ldots,\,6$ stay as they are
in~\eqref{eq:general-form-local-fr-3b-LO-pseudo-potential} while the 
matrix elements are symmetrised. 
The latter of these possibilities is adopted in the following.

We use the generalised coordinates $x_i \equiv (\bfr_i s_i q_i)$
to write the matrix elements of the interaction
in spin, isospin and coordinate spaces
\begin{equation}
    V_{3,bc}
    (x_1,x_2,x_3 ; x_4,x_5,x_6)
    = \braket{ s_1 q_1,s_2 q_2,s_3 q_3 |
    \hat{V}_{3,bc} (\bfr_1,\bfr_2,\bfr_3;\bfr_4,\bfr_5,\bfr_6)
    |s_4 q_4,s_5 q_5,s_6 q_6}\,,
\end{equation}
and we write the symmetrised matrix element as
\begin{equation}
    \bar{V}_{3,bc}
    ( x_1,x_2,x_3 ;x_4,x_5,x_6)
    = \frac{1}{6} \sum_{i=1}^{6} 
    V^{i}_{3,bc}(x_1,x_2,x_3 ;x_4,x_5,x_6 )\,,
    \label{eq:definition-symmetrised-3b-interaction}
\end{equation}
where
\begin{align}
    V^{1}_{3,bc}(x_1,x_2,x_3;x_4,x_5,x_6)
    &=V_{3,bc}(x_1,x_2,x_3;x_4,x_5,x_6) \,,
    \label{eq:matrix-elements-V-3b-1} \\
    V^{2}_{3,bc}(x_1,x_2,x_3;x_4,x_5,x_6)
    &=V_{3,bc}(x_1,x_3,x_2;x_4,x_6,x_5) \,,
    \label{eq:matrix-elements-V-3b-2} \\
    V^{3}_{3,bc}(x_1,x_2,x_3;x_4,x_5,x_6)
    &=V_{3,bc}(x_2,x_1,x_3;x_5,x_4,x_6) \,,
    \label{eq:matrix-elements-V-3b-3} \\
    V^{4}_{3,bc}(x_1,x_2,x_3;x_4,x_5,x_6)
    &=V_{3,bc}(x_2,x_3,x_1;x_5,x_6,x_4) \,,
    \label{eq:matrix-elements-V-3b-4} \\
    V^{5}_{3,bc}(x_1,x_2,x_3;x_4,x_5,x_6)
    &=V_{3,bc}(x_3,x_1,x_2;x_6,x_4,x_5) \,,
    \label{eq:matrix-elements-V-3b-5} \\
    V^{6}_{3,bc}(x_1,x_2,x_3;x_4,x_5,x_6)
    &=V_{3,bc}(x_3,x_2,x_1;x_6,x_5,x_4)\,.
    \label{eq:matrix-elements-V-3b-6}
\end{align}
With these notations, the contributions from the term $\bar{V}_{3,bc}$
to the EDF is written as
\begin{align}
    E_{3,bc} [\rho, \breve{\rho}, \breve{\rho}^{*}]
    & = \frac{1}{6} \int\!
    \mathrm{d}x_1\, \mathrm{d}x_2\,\mathrm{d}x_3\,
    \mathrm{d}x_4\, \mathrm{d}x_5\, \mathrm{d}x_6\,
    \bar{V}_{3,bc}(x_1,x_2,x_3; x_4,x_5,x_6)
    \braket{\Phi | a^{\dagger}_{x_1} a^{\dagger}_{x_2}
    a^{\dagger}_{x_3} a_{x_6} a_{x_5} a_{x_4} | \Phi}\,, \nonumber \\
    &=\int\!\rmd^3r\,{\cal E}_{3,bc}(\bfr)
    \label{eq:symmetrised-edf-3b}
\end{align}
with the integration measure defined as
\begin{equation}
\int\! \mathrm{d}x \equiv \sum_{sq} \int \mathrm{d}^{3}r\,,
\end{equation}
and where the contractions $\braket{\Phi | a^{\dagger}_{x_1} a^{\dagger}_{x_2}
    a^{\dagger}_{x_3} a_{x_6} a_{x_5} a_{x_4} | \Phi}$
can be calculated by means of the Wick's theorem and expressed in terms of the 
density matrices~\eqref{eq:decomposition-normal-density}
and~\eqref{eq:decomposition-breve-density}.

\subsection{Linearly independent terms in the functional
\label{subsec:linearly-independent-functionals}}

Before calculating the contributions to the EDF associated to the 25 Hermitian combinations 
of spin and isospin operators, it is important to keep in mind
that, 
due to the specific form of the pseudopotential considered in this work, 
further reduction of the size of the sets ${\cal I}_3^{\sigma}$ and ${\cal
I}_3^{\tau}$
is possible. Indeed, the fact that the potential is semi-regularised implies 
that the spatial exchange operator of particles 2 and 3 corresponds to the identity.
The antisymmetry of fermionic states then implies that
$\hat{P}^{\tau}_{23}$ is equivalent to $- \hat{P}^{\sigma}_{23}$ which allows to
remove $\hat{P}^{\tau}_{23}$ from the set ${\cal I}_3^{\tau}$ and leaves
$\hat{\mathbb{1}}^{\tau}$, $\hat{P}^{\tau}_{12}$ and $\hat{P}^{\tau}_{13}$ as the
only independent isospin operators.
In the remainder of the article, we therefore restrict ${\cal I}_3^{\tau}$ to these
three operators and redefined this set as
\begin{align}
    {\cal I}_3^{\tau}= \left\{
    \hat{\mathcal{P}}^{\tau}_{1} = \hat{\mathbb{1}}^{\tau}\,,~
    \hat{\mathcal{P}}^{\tau}_{2} = \hat{P}^{\tau}_{12}\,,~
    \hat{\mathcal{P}}^{\tau}_{3} = \hat{P}^{\tau}_{13}~
    \right\}\,.
    \label{eq:irreducible-ordered-set-I-tau-3b}
\end{align}
The above irreducible set along with~(\ref{eq:ordered-set-I-sigma-3b}) 
allows for generating $5 \times 3 = 15$ Hermitian 
terms that can now be written in a similar fashion as in
eq.~\eqref{eq:general-form-non-local-fr-2b-NpLO-pseudo-potential-WBHM}
for a two-body potential, namely 
\begin{align}
    \hat{V}_{3} (\bfr_1, \bfr_2, \bfr_3 ; \bfr_4, \bfr_5, \bfr_6)
    & = \sum_{\hat{\cal P}^{\sigma}_{b} \in {\cal I}_3^{\sigma}}
    \sum_{\hat{\cal P}^{\tau}_{c} \in {\cal I}_3^{\tau}}
    \hat{V}_{3,bc} (\bfr_1, \bfr_2, \bfr_3 ; \bfr_4, \bfr_5, \bfr_6)\,,
    \nonumber \\
    & = \delta (\mathbf{r}_{14}) \delta (\mathbf{r}_{25})
    \delta (\mathbf{r}_{36})
    g_{a} (\mathbf{r}_{12}) \delta (\mathbf{r}_{23})
    \sum_{\hat{\cal P}^{\sigma}_{b} \in {\cal I}_3^{\sigma}}
    \sum_{\hat{\cal P}^{\tau}_{c} \in {\cal I}_3^{\tau}}
    W_{3,bc}
    \hat{\mathcal{P}}^{\sigma}_{b}
    \hat{\mathcal{P}}^{\tau}_{c}\,.
    \label{eq:summation-irreducible-sets-LO-V-3b}
\end{align}
After deriving the 15 associated contributions to the EDF from~(\ref{eq:symmetrised-edf-3b}),
it was found that six of them are linearly dependent, namely 
$E_{3,21} = E_{3,31}\,,$
$E_{3,22} = E_{3,33}\,,$
$E_{3,32} = E_{3,23}\,,$
$E_{3,12} = E_{3,13}\,,$
$E_{3,42} = E_{3,43}\,,$ and 
$E_{3,52} = E_{3,53}\,.$
The conclusion is that the 
form~\eqref{eq:general-form-local-fr-3b-LO-pseudo-potential} 
enables one to  produce an interaction with nine independent degrees of freedom. 

%
%

\subsection{Restricted functional}

As explained in the previous section, the three-body LO semi-contact interaction
gives rise to nine independent degrees of freedom. This, of course, represents
large parameter space with a potential flexibility
to tune the functional, but also a real challenge. Indeed,
in a previous study~\cite{bennaceur_nonlocal_2017}, 
we observed that functionals build from
an interaction with two-body terms only, and containing 9 to 17 parameters,
was poorly constrained in various directions of the parameter space. 
This situation could of course be improved
if more constraints were employed to build the penalty function used to adjust the
parameters but adding nine additional parameters would anyway be very challenging.
A sensible strategy could consist in reducing the number of free parameters in
the three-body sector while preserving enough
relevant degrees of freedom.

Requiring that the functional is regular, {\it i.e.} free from the ultraviolet
divergence due to the dependence of the functional on local anomalous
densities~\cite{PhysRevLett.88.042504}, 
imposes specific relations between
the different contributions from the three-body interaction and therefore
considerably reduces the number of independent parameters. If the densities
allow for proton-neutron mixing,
it is easy to show that this is not possible
because of the $\delta$ function present in the form factor of
the three-body semi-contact interaction. The energy density will indeed contain
terms with local anomalous scalar and vector densities which 
cannot all be cancelled unless the three-body interaction is trivial, {\it i.e.}
with all strength parameters set to zero.

If we neglect the possible mixing between protons and neutrons in the
functional~\cite{perlinska_local_2004}, the anomalous local vector densities
are zero
and it is possible to construct a non-trivial interaction that leads to an energy
density that does not contain anomalous local scalar densities
$\tilde\rho_q(\bfr)$, as we will show in the following.
In addition to reducing the number of free parameters, this avoids introducing
a cut-off that regularises the ultraviolet divergence induced by
$\tilde{\rho}_{q}(\bfr)$~\cite{PhysRevLett.88.042504}.

The calculation of all terms in the energy generated by the three-body
semi-contact interaction without assumption about symmetry and with
proton-neutron mixing was done using a symbolic calculation software.

The coefficients of the terms containing at least one local pairing density
are reported on table~\ref{tab:coef-vanishing-local-pair-densities2}.
The first line gives the corresponding products of densities with the shorthand notations
\begin{align}
    & A_{\rho_{0}, \vec{\breve{\rho}}, \vec{\breve{\rho}}^{*}}
    \equiv \rho_{0} (\mathbf{r}_{\mu})
    \vec{\breve{\rho}} (\mathbf{r}_{\nu}) \circ 
    \vec{\breve{\rho}}^{*} (\mathbf{r}_{\nu})\,,
    \label{eq:notation-A-rho0-rhob-rhobs} \\
    & B_{\vec{\breve{\rho}}, \rho_{0}, \vec{\breve{\rho}}^{*}}
    \equiv \vec{\breve{\rho}} (\mathbf{r}_{\mu}) \circ 
    \rho_{0} (\mathbf{r}_{\nu}, \mathbf{r}_{\mu})
    \vec{\breve{\rho}}^{*} (\mathbf{r}_{\mu}, \mathbf{r}_{\nu})\,,
    \label{eq:notation-B-rhob-rho0-rhobs} \\
    & B_{\vec{\breve{\rho}}, \vec{\rho}, \breve{\rho}_{0}^{*}}
    \equiv \vec{\breve{\rho}} (\mathbf{r}_{\mu}) \circ 
    \vec{\rho} (\mathbf{r}_{\nu}, \mathbf{r}_{\mu})
    \breve{\rho}_{0}^{*} (\mathbf{r}_{\mu}, \mathbf{r}_{\nu})\,,
    \label{eq:notation-B-rhob-rho-rho0bs} \\
    & B_{\vec{\breve{\rho}}, \mathbf{s}_{0}, \vec{\breve{\mathbf{s}}}^{*}}
    \equiv \vec{\breve{\rho}} (\mathbf{r}_{\mu}) \circ 
    \mathbf{s}_{0} (\mathbf{r}_{\nu}, \mathbf{r}_{\mu}) \cdot 
    \vec{\breve{\mathbf{s}}}^{*} (\mathbf{r}_{\mu}, \mathbf{r}_{\nu})\,,
    \label{eq:notation-B-rhob-s0-sbs} \\
    & B_{\vec{\breve{\rho}}, \vec{\mathbf{s}}, \breve{\mathbf{s}}_{0}^{*}}
    \equiv \vec{\breve{\rho}} (\mathbf{r}_{\mu}) \circ 
    \vec{\mathbf{s}} (\mathbf{r}_{\nu}, \mathbf{r}_{\mu}) \cdot 
    \breve{\mathbf{s}}_{0}^{*} (\mathbf{r}_{\mu}, \mathbf{r}_{\nu})\,.
    \label{eq:notation-B-rhob-s-s0bs}
\end{align}
The contributions to the functional are obtained by multiplying
the coefficients from the table with the strength parameters $W_{3,bc}$ 
(or combinations of) from the same line and the products of densities
from the same column.
Note that when the Hermitian conjugate of these products of densities appears
in the functional, they are weighted with the same coefficients because of the
reality of the SR energy.

%
%

\begin{table}[htbp]
    \centering
    \caption{\label{tab:coef-vanishing-local-pair-densities2}
    Coefficients of the terms from the tree-body semi-contact interaction
    containing one anomalous local density. The contributions to the energy density defined
    in~(\ref{eq:symmetrised-edf-3b}) are obtained by multiplying these coefficients
    by the strengths parameters $W_{3,bc}$ (or com\-bi\-na\-tions of) from the
    first column and the products of densities from the first line
    with the shorthand notations given by eqs.~(\ref{eq:notation-A-rho0-rhob-rhobs})
    to~(\ref{eq:notation-B-rhob-s-s0bs}).
    Five associated complex conjugate terms are not reported since the reality
    of the energy implies that they are weighted by the same coefficients.
    On the first four lines, the terms of the EDF are constructed using the isospin
    identity operator for $c = 1$, and on the next four lines for
    $c = 2$. The case $c = 3$ does not appear because the functionals obtained 
    for this isospin exchange operator are linearly dependent from the ones 
    of the case $c=2$ (see
    section~\ref{subsec:linearly-independent-functionals}).}
    \begin{tabular}{lrrrrr}
        \toprule
        & $A_{\rho_{0}, \vec{\breve{\rho}}, \vec{\breve{\rho}}^{*}}$
        & $B_{\vec{\breve{\rho}}, \rho_{0}, \vec{\breve{\rho}}^{*}}$
        & $B_{\vec{\breve{\rho}}, \vec{\rho}, \breve{\rho}_{0}^{*}}$
        & $B_{\vec{\breve{\rho}}, \mathbf{s}_{0}, \vec{\breve{\mathbf{s}}}^{*}}$
        & $B_{\vec{\breve{\rho}}, \vec{\mathbf{s}}, \breve{\mathbf{s}}_{0}^{*}}$
        \\
        \midrule
        $W_{3,11}$
        & $\frac{1}{24}$~~ & $-\frac{1}{48}$~~ & $\frac{1}{48}$~~
        & $\frac{1}{48}$~~ & $-\frac{1}{48}$~~ \\
        \addlinespace[4pt]
        $W_{3,21}$
        & $\frac{1}{48}$~~ & $-\frac{1}{96}$~~ & $\frac{1}{96}$~~
        & $\frac{1}{96}$~~ & $-\frac{1}{96}$~~ \\
        \addlinespace[4pt]
        $W_{3,41}$
        & $-\frac{1}{24}$~~ & $\frac{1}{48}$~~ & $-\frac{1}{48}$~~
        & $-\frac{1}{48}$~~ & $\frac{1}{48}$~~ \\
        \addlinespace[4pt]
        $W_{3,51}$
        & $-\frac{1}{48}$~~ & $\frac{1}{96}$~~
        & $-\frac{1}{96}$~~ & $-\frac{1}{96}$~~ & $\frac{1}{96}$~~ \\
        \midrule
        $W_{3,12}$
        & $\frac{1}{48}$~~ & $-\frac{1}{32}$~~ & $-\frac{1}{96}$~~
        & $\frac{1}{32}$~~ & $\frac{1}{96}$~~ \\
        \addlinespace[4pt]
        $W_{3,42}$
        & $-\frac{1}{48}$~~ & $\frac{1}{32}$~~ & $\frac{1}{96}$~~
        & $-\frac{1}{32}$~~ & $-\frac{1}{96}$~~ \\
        \addlinespace[4pt]
        $W_{3,52}$
        & $-\frac{1}{96}$~~ & $\frac{1}{64}$~~
        & $\frac{1}{192}$~~ & $-\frac{1}{64}$~~ & $-\frac{1}{192}$~~ \\
        \addlinespace[4pt]
        $W_{3,22}+W_{3,32}$
        & $\frac{1}{48}$~~ & $-\frac{1}{32}$~~
        & $-\frac{1}{96}$~~ & $\frac{1}{32}$~~ & $\frac{1}{96}$~~ \\
        \bottomrule
    \end{tabular}    
\end{table}


Table~\ref{tab:coef-vanishing-local-pair-densities2} also contains the coefficient
for the term $E_{3,22}+E_{3,32}$ which is helpful to
introduce because it can easily be combined with other terms to cancel the
contributions with local anomalous scalar densities.
Additionally, we can see on
table~\ref{tab:coef-vanishing-local-pair-densities2} that it is not 
possible to make the contribution of the local pairing densities to the energy
vanish by summing one term obtained with the coupling constants $W_{3,bc}$
with $c = 1$ ({\it i.e.} from the first four lines of the table) and one 
term obtained
with $c = 2$ ({\it i.e.} from the next four lines of the table).
Therefore we build
combinations of terms corresponding to the same value of 
$c$. For $c=1$, there exist
six possibilities to
cancel the contributions with anomalous scalar densities, namely
\begin{align}
E_1&=E_{3,11}+E_{3,41}\,,     \label{eq:combination-E1} \\
E_2&=E_{3,21}+E_{3,51}\,,     \label{eq:combination-E2} \\
E_3&=E_{3,11}+2E_{3,51}\,,    \label{eq:combination-E3} \\
E_4&=E_{3,11}-2E_{3,21}\,,    \label{eq:combination-E4} \\
E_5&=E_{3,41}-2E_{3,51}\,,    \label{eq:combination-E5} \\
E_6&=2E_{3,21}+E_{3,41}\,.    \label{eq:combination-E6}
\end{align}
For $c=2$, there also exist six possibilities to
cancel the contributions with anomalous scalar densities which are
\begin{align}
E_1'&=E_{3,12}+E_{3,42}\,, \\
E_2'&=E_{3,52}+\tfrac{1}{2}\left(E_{3,22}+E_{3,32}\right)\,, \\
E_3'&=E_{3,12}+2E_{3,52}\,, \\
E_4'&=E_{3,12}-E_{3,22}-E_{3,32}\,, \\
E_5'&=E_{3,42}-2E_{3,52}\,, \\
E_6'&=E_{3,42}+E_{3,22}+E_{3,32}\,.
\end{align}

Further simplifications can be achieved by noticing that  $E_{n} = 2 E_{n}'$
for $n \in \{ 1, \ldots, 6 \}$\,,
which implies that we can, for instance, consider only the contributions to the
functional~\eqref{eq:combination-E1} to~\eqref{eq:combination-E6}.
This choice is motivated by the fact that the effective interactions 
generating these six contributions to the EDF are diagonal in isospin ({\it i.e.}
with $c=1$).
Moreover, writing out the full expressions for $E_3$, $E_4$, $E_5$, and $E_6$, we found that $E_3 = E_6$ and 
$E_4 = E_5$\,, 
which eliminates two further functionals, {\it e.g.} $E_5$ and $E_6$, from the set of terms that have to be considered. Finally, the two relations
\begin{equation}
    E_3 - E_4 = 2 E_2\,,
    \label{eq:constraint-eq-1-functional}
\end{equation}
and
\begin{equation}
    E_4 + E_6 = E_1\,,
    \label{eq:constraint-eq-2-functional}
\end{equation}
show that only two linearly independent contributions to the EDF make the local pairing
densities vanish. We kept $E_1$ that corresponds exactly to the
functional derived in~\cite{costa_interactions_2022} as well as
$E_2$\,. With these choices, the final form of the semi-regularised pseudopotential reads 
\begin{align}
    \hat{V}_{3} (\bfr_1, \bfr_2, \bfr_3; \bfr_4, \bfr_5, \bfr_6)
    &= \delta (\mathbf{r}_{14}) \delta (\mathbf{r}_{25})
    \delta (\mathbf{r}_{36})
    \biggl\{
    W_{3,1} \left( \hat{\mathbb{1}}^{\sigma} + \hat{P}^{\sigma}_{23} \right)\nonumber \\
    &\hskip 1cm
    + W_{3,2} \left[
    \frac{1}{2}\left(\hat{P}^{\sigma}_{12} \hat{P}^{\sigma}_{23}
    + \hat{P}^{\sigma}_{23} \hat{P}^{\sigma}_{12}\right)
    + \hat{P}^{\sigma}_{12}
    \right]    \biggr\}
    \hat{\mathbb{1}}^{\tau}
    g_{a} (\mathbf{r}_{12}) \delta (\mathbf{r}_{23})\,,
    \label{eq:final-form-V-3b}
\end{align}
where the two real parameters $W_{3,1}$ and $W_{3,2}$ represent the strength of the two terms.
The latter are expressed with the strengths of the individual terms
from~(\ref{eq:summation-irreducible-sets-LO-V-3b}) as 
\begin{align}
    W_{3,1} &= W_{3,11} = W_{3,41}\,,
    \label{eq:W1-as-function-of-W-beta-gamma} \\
    W_{3,2}
    &= W_{3,51}
    = W_{3,21}\,.
    \label{eq:W2-as-function-of-W-beta-gamma}
\end{align}
We recall that the indices 1, 2 and 3 in
equation~\eqref{eq:final-form-V-3b} do \textit{not} refer to the indices
of the spatial  coordinates $\mathbf{r}_{1},$ $\mathbf{r}_{2}$,
and $\mathbf{r}_{3}$ (or spin coordinates
$s_{1}$, $s_{2}$, and $s_{3}$) but on the positions of the spatial or spin coordinates in the three-body ket on which the operators act.
Proper symmetrisation will be ensured by the use of 
$\bar{V}_{3,bc}$~\eqref{eq:definition-symmetrised-3b-interaction} when 
calculating the EDF~\eqref{eq:symmetrised-edf-3b}.

\subsection{Energy density functional\label{subsec:energy-density-functional}}

Evaluation of~\eqref{eq:symmetrised-edf-3b} for the  
pseudopotential~\eqref{eq:final-form-V-3b} yields the trilinear
particle-hole and particle-particle parts of the EDF. Their most
general expressions, with possible proton-neutron mixing and
using the ``breve''
representation, are
given in appendix~\ref{app:functional-with-pn}. In this paragraph,
we give their forms when proton-neutron mixing is disregarded within
the more widely-used ``tilde'' representation. The trilinear
particle-hole (ph) part of the functional then reads
\begin{align}
    E_{3,\text{ph}}
    = \int\! \mathrm{d}^{3}r_{1} \mathrm{d}^{3}r_{2}\,&g_{a} (\mathbf{r}_{12})
    \Big\{
    A^{\rho_{0}}_{\rho_{0} \rho_{0}}
    \rho_{0} (\mathbf{r}_{1}) \rho_{0}^{2} (\mathbf{r}_{2})
    + A^{\rho_{0}}_{\rho_{1} \rho_{1}}
    \rho_{0} (\mathbf{r}_{1}) \rho_{1}^{2} (\mathbf{r}_{2})
    + A^{\rho_{0}}_{\mathbf{s}_{0} \mathbf{s}_{0}}
    \rho_{0} (\mathbf{r}_{1}) \mathbf{s}_{0}^{2} (\mathbf{r}_{2})
    \nonumber \\
    & + A^{\rho_{0}}_{\mathbf{s}_{1} \mathbf{s}_{1}}
    \rho_{0} (\mathbf{r}_{1}) \mathbf{s}_{1}^{2} (\mathbf{r}_{2})
    + A^{\mathbf{s}_{0}}_{\rho_{0} \mathbf{s}_{0}}
    \rho_{0} (\mathbf{r}_{1}) \mathbf{s}_{0} (\mathbf{r}_{1})
    \cdot \mathbf{s}_{0} (\mathbf{r}_{2})
    + A^{\mathbf{s}_{0}}_{\rho_{1} \mathbf{s}_{1}}
    \rho_{1} (\mathbf{r}_{1}) \mathbf{s}_{1} (\mathbf{r}_{1})
    \cdot \mathbf{s}_{0} (\mathbf{r}_{2})
    \nonumber \\
    & + \frac{1}{2} \big[ \rho_{0} (\mathbf{r}_{1})
    + \rho_{0} (\mathbf{r}_{2}) \big] \big[
    B^{\rho_{0}}_{\rho_{0} \rho_{0}}
    \rho_{0} (\mathbf{r}_{1}, \mathbf{r}_{2})
    \rho_{0} (\mathbf{r}_{2}, \mathbf{r}_{1})
    + B^{\rho_{0}}_{\rho_{1} \rho_{1}}
    \rho_{1} (\mathbf{r}_{1}, \mathbf{r}_{2})
    \rho_{1} (\mathbf{r}_{2}, \mathbf{r}_{1})
    \nonumber \\
    & \hspace{3cm} + B^{\rho_{0}}_{\mathbf{s}_{0} \mathbf{s}_{0}}
    \mathbf{s}_{0} (\mathbf{r}_{1}, \mathbf{r}_{2}) \cdot
    \mathbf{s}_{0} (\mathbf{r}_{2}, \mathbf{r}_{1})
    + B^{\rho_{0}}_{\mathbf{s}_{1} \mathbf{s}_{1}}
    \mathbf{s}_{1} (\mathbf{r}_{1}, \mathbf{r}_{2}) \cdot
    \mathbf{s}_{1} (\mathbf{r}_{2}, \mathbf{r}_{1})
    \big]
    \nonumber \\
    & + \frac{1}{2} \big[ \rho_{1} (\mathbf{r}_{1})
    + \rho_{1} (\mathbf{r}_{2}) \big] \big[
    B^{\rho_{1}}_{\rho_{0} \rho_{1}}
    \rho_{0} (\mathbf{r}_{1}, \mathbf{r}_{2})
    \rho_{1} (\mathbf{r}_{2}, \mathbf{r}_{1})
    + B^{\rho_{1}}_{\mathbf{s}_{0}\mathbf{s}_{1}}
    \mathbf{s}_{0} (\mathbf{r}_{1}, \mathbf{r}_{2}) \cdot 
    \mathbf{s}_{1} (\mathbf{r}_{2}, \mathbf{r}_{1})
    \big]
    \nonumber \\
    & + \frac{1}{2} \big[ \mathbf{s}_{0} (\mathbf{r}_{1})
    + \mathbf{s}_{0} (\mathbf{r}_{2}) \big] \cdot \big[
    B^{\mathbf{s}_{0}}_{\rho_{0} \mathbf{s}_{0}}
    \rho_{0} (\mathbf{r}_{1}, \mathbf{r}_{2})
    \mathbf{s}_{0} (\mathbf{r}_{2}, \mathbf{r}_{1})
    + B^{\mathbf{s}_{0}}_{\rho_{1} \mathbf{s}_{1}}
    \rho_{1} (\mathbf{r}_{1}, \mathbf{r}_{2})
    \mathbf{s}_{1} (\mathbf{r}_{2}, \mathbf{r}_{1})
    \big]
    \nonumber \\
    & + \frac{1}{2}
    \big[ \mathbf{s}_{1} (\mathbf{r}_{1})
    + \mathbf{s}_{1} (\mathbf{r}_{2}) \big] \cdot \big[
    B^{\mathbf{s}_{1}}_{\rho_{0} \mathbf{s}_{1}}
    \rho_{0} (\mathbf{r}_{1}, \mathbf{r}_{2})
    \mathbf{s}_{1} (\mathbf{r}_{2}, \mathbf{r}_{1})
    + B^{\mathbf{s}_{1}}_{\rho_{1} \mathbf{s}_{0}}
    \rho_{1} (\mathbf{r}_{1}, \mathbf{r}_{2})
    \mathbf{s}_{0} (\mathbf{r}_{2}, \mathbf{r}_{1})
    \big]
    \nonumber \\
    & + \mathrm{i}\, \mathbf{s}_{0} (\mathbf{r}_{1}) \cdot \big[
    T^{\mathbf{s}_{0}}_{\mathbf{s}_{0}\mathbf{s}_{0}}
    \mathbf{s}_{0} (\mathbf{r}_{1}, \mathbf{r}_{2}) \times
    \mathbf{s}_{0} (\mathbf{r}_{2}, \mathbf{r}_{1})
    + T^{\mathbf{s}_{0}}_{\mathbf{s}_{1}\mathbf{s}_{1}}
    \mathbf{s}_{1} (\mathbf{r}_{1}, \mathbf{r}_{2}) \times
    \mathbf{s}_{1} (\mathbf{r}_{2}, \mathbf{r}_{1})
    \big]
    \nonumber \\
    & + \mathrm{i}\, \big[
    \mathbf{s}_{1} (\mathbf{r}_{1}) - \mathbf{s}_{1} (\mathbf{r}_{2})
    \big] \cdot \big[
    T^{\mathbf{s}_{1}}_{\mathbf{s}_{0}\mathbf{s}_{1}}
    \mathbf{s}_{0} (\mathbf{r}_{1}, \mathbf{r}_{2}) \times
    \mathbf{s}_{1} (\mathbf{r}_{2}, \mathbf{r}_{1})
    \big]
    \Big\}\,,
    \label{eq:E-3b-ph-no-pn-with-cc}
\end{align}
where $\rho_1$ and $\mathbf s_1$ are given by
equations~(\ref{eq:definition-scalar-isovector-density-without-pn-mixing})
and~(\ref{eq:definition-vector-isovector-density-without-pn-mixing}).

For the particle-particle (pp) part, one has 
\begin{align}
    E_{3,\text{pp}}
    = \frac{1}{2} \sum_{q = \text{n, p}}
    \int \mathrm{d}^{3}&r_{1}\, \mathrm{d}^{3}r_{2}\,g_{a} (\mathbf{r}_{12}) \nonumber \\
    & \times\Big\{ \left[ \rho_{\bar{q}} (\mathbf{r}_{1})
    + \rho_{\bar{q}} (\mathbf{r}_{2}) \right] \left[
    C^{\rho}_{\tilde{\rho} \tilde{\rho}}
    \tilde{\rho}_{q} (\mathbf{r}_{1}, \mathbf{r}_{2})
    \tilde{\rho}_{q}^{*} (\mathbf{r}_{1}, \mathbf{r}_{2})
    + C^{\rho}_{\tilde{\mathbf{s}} \tilde{\mathbf{s}}}
    \tilde{\mathbf{s}}_{q} (\mathbf{r}_{1}, \mathbf{r}_{2}) \cdot 
    \tilde{\mathbf{s}}_{q}^{*} (\mathbf{r}_{1}, \mathbf{r}_{2})
    \right]
    \nonumber \\
    & + \left[ \mathbf{s}_{\bar{q}} (\mathbf{r}_{1})
    - \mathbf{s}_{\bar{q}} (\mathbf{r}_{2}) \right] \cdot \left[
    C^{\mathbf{s}}_{\tilde{\rho} \tilde{\mathbf{s}}}
    \tilde{\rho}_{q} (\mathbf{r}_{1}, \mathbf{r}_{2})
    \tilde{\mathbf{s}}_{q}^{*} (\mathbf{r}_{1}, \mathbf{r}_{2})
    + C^{\mathbf{s}}_{\tilde{\mathbf{s}} \tilde{\rho}}
    \tilde{\mathbf{s}}_{q} (\mathbf{r}_{1}, \mathbf{r}_{2})
    \tilde{\rho}_{q}^{*} (\mathbf{r}_{1}, \mathbf{r}_{2})
    \right]
    \nonumber \\
    & + \mathrm{i}\, C^{\mathbf{s}}_{\tilde{\mathbf{s}}\tilde{\mathbf{s}}} \left[
    \mathbf{s}_{\bar{q}} (\mathbf{r}_{1})
    + \mathbf{s}_{\bar{q}} (\mathbf{r}_{2})
    \right] \cdot \left[
    \tilde{\mathbf{s}}_{q} (\mathbf{r}_{1}, \mathbf{r}_{2}) \times
    \tilde{\mathbf{s}}^{*}_{q} (\mathbf{r}_{1}, \mathbf{r}_{2})
    \right]
    \Big\}\,.
    \label{eq:E-3b-pp-no-pn-with-cc}
\end{align}
The expressions for
the coupling constants, denoted with letters $A$, $B$, $T$
and $C$, can be found in 
appendix~\ref{app:coupling-constants}.
For the particle-hole channel, they obey the following convention:
letter $A$ weights
terms involving three local densities, letter $B$ terms with non-local
densities, one being scalar and the two other ones being either scalars or vectors,
letter $T$ weights terms with three vector densities (two of which being non-local).
Coupling constants for the particle-particle part of the functional are
all denoted with letter $C$. For both channels, the density in superscript
corresponds to a local density and the ones in subscript to two other
densities, either local or non-local, coupled to the first one to form a
scalar-isoscalar term.
As explained at the beginning of this section, the two
contributions given by equations~(\ref{eq:E-3b-ph-no-pn-with-cc}) and~(\ref{eq:E-3b-pp-no-pn-with-cc}) correspond to an EDF calculated without
proton-neutron mixing (expressions with
proton-neutron mixing are given in 
appendix~\ref{app:functional-with-pn}). As it can be seen from
the prescriptions
from table~\ref{tab:densities-from-tilde-to-breve-representation}, isovector
densities $\vec \rho$ and $\vec{\mathbf s}$ have only their third components
(as is customary, denoted $\rho_1$ and $\mathbf s_1$) possibly non-zero, and
the particle-particle part depends on proton and neutrons anomalous
densities.


For a time-reversal-invariant
system, all terms that contain local spin densities vanish and
therefore one has
\begin{align}
    E_{3,\text{ph}}^{\text{time-even}}
    & = \int \mathrm{d}^{3}r_{1} \mathrm{d}^{3}r_{2}~g_{a} (\mathbf{r}_{12})
    \Bigl\{
    \left(
    \tfrac{1}{8}\, W_{3,1} + \tfrac{1}{16}\, W_{3,2}
    \right)
    \rho_{0} (\mathbf{r}_{1}) \left[
    \rho_{0}^{2} (\mathbf{r}_{2}) - \rho_{1}^{2} (\mathbf{r}_{2})
    \right]
    \nonumber \\
    & - \frac{1}{2} \left[ \rho_{0} (\mathbf{r}_{1})
    + \rho_{0} (\mathbf{r}_{2}) \right] \left(
    \tfrac{1}{16}\, W_{3,1} + \tfrac{3}{32}\, W_{3,2}
    \right) \left[
    \rho_{0} (\mathbf{r}_{1}, \mathbf{r}_{2})
    \rho_{0} (\mathbf{r}_{2}, \mathbf{r}_{1})
    + \rho_{1} (\mathbf{r}_{1}, \mathbf{r}_{2})
    \rho_{1} (\mathbf{r}_{2}, \mathbf{r}_{1})
    \right]
    \nonumber \\
    & - \frac{1}{2} \left[ \rho_{0} (\mathbf{r}_{1})
    + \rho_{0} (\mathbf{r}_{2}) \right] \left(
    \tfrac{1}{16}\, W_{3,1} + \tfrac{1}{96}\, W_{3,2}
    \right) \left[
    \mathbf{s}_{0} (\mathbf{r}_{1}, \mathbf{r}_{2}) \cdot
    \mathbf{s}_{0} (\mathbf{r}_{2}, \mathbf{r}_{1})
    + \mathbf{s}_{1} (\mathbf{r}_{1}, \mathbf{r}_{2}) \cdot
    \mathbf{s}_{1} (\mathbf{r}_{2}, \mathbf{r}_{1})
    \right]
    \nonumber \\
    & + \frac{1}{2} \left[ \rho_{1} (\mathbf{r}_{1})
    + \rho_{1} (\mathbf{r}_{2}) \right] \left(
    \tfrac{1}{8}\, W_{3,1} + \tfrac{3}{16}\, W_{3,2}
    \right)
    \rho_{0} (\mathbf{r}_{1}, \mathbf{r}_{2})
    \rho_{1} (\mathbf{r}_{2}, \mathbf{r}_{1})
    \nonumber \\
    & + \frac{1}{2} \left[ \rho_{1} (\mathbf{r}_{1})
    + \rho_{1} (\mathbf{r}_{2}) \right] \left(
    \tfrac{1}{8}\, W_{3,1} + \tfrac{1}{48}\, W_{3,2}
    \right)
    \mathbf{s}_{0} (\mathbf{r}_{1}, \mathbf{r}_{2}) \cdot 
    \mathbf{s}_{1} (\mathbf{r}_{2}, \mathbf{r}_{1})
    \Bigr\}\,,
    \label{eq:E-3b-ph-no-pn-time-even}
\end{align}
and
\begin{align}
    E_{3,\text{pp}}^{\text{time-even}}
    &= \frac{1}{2} \sum_{q = \text{n, p}}
    \int\! \mathrm{d}^{3}r_{1}\, \mathrm{d}^{3}r_{2}\,g_{a} (\mathbf{r}_{12})
    \left[ \rho_{\bar{q}} (\mathbf{r}_{1})
    + \rho_{\bar{q}} (\mathbf{r}_{2}) \right]
    \nonumber \\
    & \times  \left[ \left(
    \tfrac{1}{4}\, W_{3,1} - \tfrac{1}{8}\, W_{3,2}
    \right)
    \tilde{\rho}_{q} (\mathbf{r}_{1}, \mathbf{r}_{2})
    \tilde{\rho}_{q}^{*} (\mathbf{r}_{1}, \mathbf{r}_{2})
    + \left(
    \tfrac{1}{4}\, W_{3,1} + \tfrac{5}{24}\, W_{3,2}
    \right)
    \tilde{\mathbf{s}}_{q} (\mathbf{r}_{1}, \mathbf{r}_{2}) \cdot 
    \tilde{\mathbf{s}}_{q}^{*} (\mathbf{r}_{1}, \mathbf{r}_{2})
    \right]\,.
    \label{eq:E-3b-pp-no-pn-time-even}
\end{align}
Equations~\eqref{eq:E-3b-ph-no-pn-time-even} and~\eqref{eq:E-3b-pp-no-pn-time-even}
are written using the parameters of the interaction given
by~\eqref{eq:W1-as-function-of-W-beta-gamma} and~\eqref{eq:W2-as-function-of-W-beta-gamma}
instead of the coupling constants
as written in appendix~\ref{app:coupling-constants} in order to highlight the fact
that the three-body interaction in its present form contains two independent
parameters usable to tune 
the functional in the ph and pp channels, and
to make more transparent the discussion in the next section.

\subsection{The RegMR3 parametrisation\label{sec:regmr3-parametrization}}

A special case of the semi-contact three-body interaction
was already considered earlier in an exploratory work~\cite{costa_interactions_2022}.
This version, used on top of a regularised N3LO interaction, corresponds
to the special choice for which $W_{3,2}=0$ and therefore three-body contributions
to the functional for time-even nuclei which are
\begin{align}
    E_{3,\text{ph}}^{\text{time-even}}
    & = \frac{1}{8}W_{3,1}
    \int \mathrm{d}^{3}r_{1} \mathrm{d}^{3}r_{2}~g_{a} (\mathbf{r}_{12})
    \Bigl\{
    \rho_{0} (\mathbf{r}_{1}) \left[
    \rho_{0}^{2} (\mathbf{r}_{2}) - \rho_{1}^{2} (\mathbf{r}_{2})
    \right]
    \nonumber \\
    & - \frac{1}{4} \left[ \rho_{0} (\mathbf{r}_{1})
    + \rho_{0} (\mathbf{r}_{2}) \right]
    \left[
    \rho_{0} (\mathbf{r}_{1}, \mathbf{r}_{2})
    \rho_{0} (\mathbf{r}_{2}, \mathbf{r}_{1})
    + \rho_{1} (\mathbf{r}_{1}, \mathbf{r}_{2})
    \rho_{1} (\mathbf{r}_{2}, \mathbf{r}_{1}) \right. \nonumber \\
    &\hskip 4cm \left.
    +\mathbf{s}_{0} (\mathbf{r}_{1}, \mathbf{r}_{2}) \cdot
    \mathbf{s}_{0} (\mathbf{r}_{2}, \mathbf{r}_{1})
    + \mathbf{s}_{1} (\mathbf{r}_{1}, \mathbf{r}_{2}) \cdot
    \mathbf{s}_{1} (\mathbf{r}_{2}, \mathbf{r}_{1})
    \right]
    \nonumber \\
    &
    + \frac{1}{2} \left[ \rho_{1} (\mathbf{r}_{1})
    + \rho_{1} (\mathbf{r}_{2}) \right]
    \left[
    \rho_{0} (\mathbf{r}_{1}, \mathbf{r}_{2})
    \rho_{1} (\mathbf{r}_{2}, \mathbf{r}_{1})
    +    \mathbf{s}_{0} (\mathbf{r}_{1}, \mathbf{r}_{2}) \cdot 
    \mathbf{s}_{1} (\mathbf{r}_{2}, \mathbf{r}_{1})\right]
    \Bigr\}\,,
    \label{eq:E-3b-ph-no-pn-time-even-regmr3}
\end{align}
and
\begin{align}
    E_{3,\text{pp}}^{\text{time-even}}
    &= \frac{1}{8}\,W_{3,1} \sum_{q = \text{n, p}}
    \int\! \mathrm{d}^{3}r_{1}\, \mathrm{d}^{3}r_{2}\,g_{a} (\mathbf{r}_{12})
    \left[ \rho_{\bar{q}} (\mathbf{r}_{1})
      + \rho_{\bar{q}} (\mathbf{r}_{2}) \right] \nonumber \\
    &\hskip 2cm\times
    \left[
    \tilde{\rho}_{q} (\mathbf{r}_{1}, \mathbf{r}_{2})
    \tilde{\rho}_{q}^{*} (\mathbf{r}_{1}, \mathbf{r}_{2})
    +
    \tilde{\mathbf{s}}_{q} (\mathbf{r}_{1}, \mathbf{r}_{2}) \cdot 
    \tilde{\mathbf{s}}_{q}^{*} (\mathbf{r}_{1}, \mathbf{r}_{2})
    \right]\,.
    \label{eq:E-3b-pp-no-pn-time-even-regmr3}
\end{align}
Although this simpler three-body interaction leads to promising results
when applied to the calculation of binding energies of nuclei close
to the valley of stability, it appeared that it is flawed by
unrealistically strong pairing correlations in very neutron-rich
nuclei and in infinite neutron matter. This can easily be understood by
considering the fact that $W_{3,1}$ must be positive for the three-body interaction
to be repulsive in the particle-hole channel and therefore
avoid the collapse of the equation of state for symmetric matter at
high density. The global effect from the
three-body interaction on pairing correlations in nuclei with $N\simeq Z$
can be inferred from the
expression~(\ref{eq:E-3b-pp-no-pn-time-even-regmr3}) written
in the case of a symmetric spin-saturated system with $N=Z$
\begin{align}
    E_{3,\text{pp}}^{\text{sym}}
    &\simeq \frac{1}{16}W_{3,1}
    \int\! \mathrm{d}^{3}r_{1}\, \mathrm{d}^{3}r_{2}\,g_{a} (\mathbf{r}_{12})
    \left[ \rho_0 (\mathbf{r}_{1})
    + \rho_0 (\mathbf{r}_{2}) \right]
    \sum_{q = \text{n, p}}
    \tilde{\rho}_{q} (\mathbf{r}_{1}, \mathbf{r}_{2})
    \tilde{\rho}_{q}^{*} (\mathbf{r}_{1}, \mathbf{r}_{2})\,.
    \label{eq:sym-pp-regmr3}
\end{align}
Since the coefficient $W_{3,1}$ has
to be positive, the corresponding term in the interaction is repulsive in the pairing channel
as it is directly seen from the previous equation. For pairing correlations
to be realistic in $N\simeq Z$ nuclei, the repulsive three-body term must
be compensated by a strong enough two-body interaction. This two-body
interaction will be attractive as well in very asymmetric systems while
the intensity of the three-body will decrease as it can be seen from
equation~(\ref{eq:E-3b-pp-no-pn-time-even-regmr3}).
As a result, average pairing gaps will be unrealistically large in very
neutron rich nuclei, neutron matter or
neutron droplets~\cite{costa_interactions_2022}.

It is enlightening to check that the parameter $W_{3,2}$ defined
in paragraph~\ref{subsec:energy-density-functional} allows for avoiding
this unwanted behaviour.
First, it can be seen from equation~(\ref{eq:E-3b-pp-no-pn-time-even})
that the contribution to pairing from the scalar anomalous density can
be tuned with the $W_{3,2}$ parameter.
Then, and even though this looks like a rather drastic choice,
it is even possible to make this term vanish by requiring $W_{3,2} = 2 W_{3,1}$. 
For a time-even system, this condition yields a functional with
\begin{align}
    & E_{3,\text{pp}}^{\text{time-even}}
    = \frac{1}{3}W_{3,1} \sum_{q = \text{n, p}}
    \int\! \mathrm{d}^{3}r_{1}\, \mathrm{d}^{3}r_{2}\,g_{a} (\mathbf{r}_{12})
    \left[
    \rho_{\bar{q}} (\mathbf{r}_{1})
    + \rho_{\bar{q}} (\mathbf{r}_{2}) \right]
    \tilde{\mathbf{s}}_{q} (\mathbf{r}_{1}, \mathbf{r}_{2}) \cdot 
    \tilde{\mathbf{s}}_{q}^{*} (\mathbf{r}_{1}, \mathbf{r}_{2})\,.
    \label{eq:E-3b-pp-no-pn-time-even-W2-equal-2W1}
\end{align}
In this limiting case, we see that the three-body interaction 
acts in the pairing channel only through the non-local anomalous vector density
$\tilde{\mathbf{s}}_q$, and 
therefore can be expected to give a small contribution only.


\section{Relations in homogeneous infinite nuclear matter\label{sec:relations-INM}}
\subsection{General properties\label{subsec:general-properties-INM}}

In this section, we provide the analytical expressions 
for the most common quantities used to characterise the properties of
homogeneous infinite nuclear matter.
Only the contributions of the three-body part of the pseudopotential will 
be given.
These contributions have to be added to the
ones of the two-body interaction that the three-body interaction will be 
combined with. We plan to use it with a regularised two-body interaction,
for which the expressions can be found in~\cite{bennaceur_new_2014}.
The auxiliary functions used throughout this section are all defined in 
appendix~\ref{app:auxiliary-functions}. A very useful property of these
functions $F_0$, $G_0$, $H_0$, and $K_0$, from which the results for a zero-range three-body
can be obtained, is that they are all equal to 1 when their argument is 0.

The following notations are used throughout this section
\begin{align}
  \xi&=ak\,, \\
  \xi_{\text{F}}&=ak_{\text{F}}\,, \\
  \xi_\mathrm{sat}&=ak_{\text{F},\mathrm{sat}}\,,
\end{align}
where $a$ is the range of the finite-range part of the form
factor from equation~(\ref{eq:normalized-gaussian-ff}), $k$ is the norm of
the momentum of a nucleon propagating in homogeneous infinite nuclear matter,
$k_\mathrm{F}$ the Fermi momentum, and $k_{F,\mathrm{sat}}$ the Fermi momentum at saturation
density.


\subsubsection{Equation of state\label{subsubsec:eos}}

In symmetric nuclear matter where $\rho_{0} \ne 0$ and
$\rho_{1} = s_{0} = s_{1} = 0$, the equation of state (EOS) is 
\begin{align}
    \frac{E}{A}
    \leftarrow \rho_{0}^{2} \big[ A^{\rho_{0}}_{\rho_{0} \rho_{0}}
    + B^{\rho_{0}}_{\rho_{0} \rho_{0}} F_{0} \left(\xi_{\text{F}}\right)
    \big]\,,
    \label{eq:eos-SNM}
\end{align}
where the symbol "$\leftarrow$" means that this expression is the
contribution from the three-body interaction to the energy per particle
in symmetric infinite nuclear matter.

In pure neutron matter where $\rho_{0} = \rho_{1} \ne 0$
and $s_{0} = s_{1} = 0$, the contribution to the EOS vanishes by construction
because the requirement to cancel terms depending on local pairing scalar densities
also cancels the contribution from the three-body interaction to neutron matter.
In polarised symmetric nuclear matter
where $\rho_{0} = s_{0} \ne 0$ and $\rho_{1} = s_{1} = 0$, 
we have
\begin{align}
    \frac{E}{A}
    \leftarrow \rho_{0}^{2} \big[
    A^{\rho_{0}}_{\rho_{0} \rho_{0}}
    + A^{\rho_{0}}_{\mathbf{s}_{0} \mathbf{s}_{0}}
    + A^{\mathbf{s}_{0}}_{\rho_{0} \mathbf{s}_{0}}
    + \big(
    B^{\rho_{0}}_{\rho_{0} \rho_{0}}
    + B^{\rho_{0}}_{\mathbf{s}_{0} \mathbf{s}_{0}}
    + B^{\mathbf{s}_{0}}_{\rho_{0} \mathbf{s}_{0}}
    \big) F_{0}\left(\xi_{\text{F}}\right)
    \big]\,.
    \label{eq:eos-polarized-SNM}
\end{align}
In fully polarised neutron matter where $\rho_{0} = \rho_{1} = s_{0} = s_{1} \ne 0$, the
contribution to the
EOS is again zero because the Pauli principle forbids the zero-range part of the three-body interaction to contribute.

Finally, for spin-isospin polarised matter where
$\rho_{0} = s_{1} \ne 0$ and $\rho_{1} = s_{0} = 0$, we have 
\begin{align}
    \frac{E}{A}
    \leftarrow \rho_{0}^{2} \big[
    A^{\rho_{0}}_{\rho_{0} \rho_{0}}
    + A^{\rho_{0}}_{\mathbf{s}_{1} \mathbf{s}_{1}}
    + \big(
    B^{\rho_{0}}_{\rho_{0} \rho_{0}}
    + B^{\rho_{0}}_{\mathbf{s}_{1} \mathbf{s}_{1}}
    + B^{\mathbf{s}_{1}}_{\rho_{0} \mathbf{s}_{1}}
    \big) F_{0}\left(\xi_{\text{F}}\right)
    \big]\,.
    \label{eq:eos-gamow-teller}
\end{align}

%
%

\subsubsection{Effective mass\label{subsubsec:effective-mass}}

In symmetric nuclear matter, the contribution of the three-body interaction
to the single-particle energy reads 
\begin{align}
    \varepsilon_{k}
    &\leftarrow 3 A^{\rho_{0}}_{\rho_{0} \rho_{0}} \rho_{0}^{2}
    + B^{\rho_{0}}_{\rho_{0} \rho_{0}} \rho_{0}^{2} F_{0}\left(\xi_{\text{F}}\right)
    \nonumber \\
    &\phantom{\leftarrow}+ \frac{2 \rho_{0}^{2}}{\xi_{\text{F}}^{3}}
    B^{\rho_{0}}_{\rho_{0} \rho_{0}} \left\{
    \frac{6}{\xi} \left[
    \mathrm{e}^{-\frac{1}{4} (\xi + \xi_{\text{F}})^{2}}
    - \mathrm{e}^{-\frac{1}{4} (\xi - \xi_{\text{F}})^{2}}
    \right]
    + 3 \sqrt{\pi} \left[
    \Erf\left(\frac{\xi + \xi_{\text{F}}}{2} \right)
    - \Erf\left(\frac{\xi-\xi_{\text{F}}}{2}\right)
    \right]
    \right\}\,.
    \label{eq:spe-SNM}
\end{align}
We can then calculate
\begin{align}
    \frac{1}{2k}\, \frac{\partial \varepsilon_{k}}{\partial k}
    \leftarrow \frac{3a^2 \rho_{0}^{2}}{\xi^{3} \xi_{\text{F}}^{3}}
    B^{\rho_{0}}_{\rho_{0} \rho_{0}} \left\{
    2 \left[ \mathrm{e}^{-\frac{(\xi - \xi_{\text{F}})^{2}}{4} }
    - \mathrm{e}^{-\frac{(\xi + \xi_{\text{F}})^{2}}{4} } \right]
    - \xi \xi_{\text{F}}
    \left[ \mathrm{e}^{-\frac{(\xi - \xi_{\text{F}})^{2}}{4} }
    + \mathrm{e}^{-\frac{(\xi + \xi_{\text{F}})^{2}}{4} } \right]
    \right\}\,.
    \label{eq:dispersion-relation-SNM}
\end{align}
This expression yields a relation that defines the contribution
to the effective mass at density $\rho_0$ and for $k=k_{\text{F}}$ through
\begin{align}
    \frac{\hbar^{2}}{2 m^{*} (\rho_{0})}
    \leftarrow - \frac{a^{2}}{2} \rho_{0}^{2} G_{0} (\xi_{\text{F}})\,.
    \label{eq:m-star-SNM}
\end{align}

%
%

\subsubsection{Pressure\label{subsubsec:pressure}}

The contribution to the pressure is obtained from the EOS in symmetric nuclear matter as
\begin{align}
    P (\rho_{0}) = \rho_{0}^{2} \frac{\partial (E/A)}{\partial \rho_{0}}
    \leftarrow 2 \rho_{0}^{3} \big[ A^{\rho_{0}}_{\rho_{0} \rho_{0}}
    + B^{\rho_{0}}_{\rho_{0} \rho_{0}} F_{0} (\xi_{\text{F}}) \big]
    + \frac{\xi_{\text{F}}}{3}\, \rho_{0}^{3}
    B^{\rho_{0}}_{\rho_{0} \rho_{0}} F_{0}' (\xi_{\text{F}})\,,
    \label{eq:contribution-pression-3b-SNM-no-polar-RegMR3}
\end{align}
where $F_0'$ is the derivative of $F_0$.
The saturation density is then defined as the non-trivial solution of 
$P (\rho_{\text{sat}}) = 0$.

%
%

\subsubsection{Compression modulus\label{subsubsec:incompressibility-modulus}}

The compression modulus expresses the energy cost necessary to
compress the nuclear fluid~\cite{BLAIZOT1980171}. It is defined as
\begin{align}
    K (\rho_{0})
    =  \frac{18 P (\rho_{0})}{\rho_{0}}
    + 9 \rho_{0}^{2} \frac{\partial^{2} (E/A)}{\partial \rho_{0}^{2}}\,,
    \label{eq:definition-incompressibility-modulus}
\end{align}
such that at equilibrium where the pressure vanishes, one has
\begin{align}
    K_{\infty}
    & = 9 \rho_{0}^{2} \frac{\partial^{2} (E/A)}{\partial \rho_{0}^{2}}
    \Bigg|_{\rho_{0} = \rho_{\text{sat}}}\,,
    \nonumber \\
    & \leftarrow 18 \rho_{\text{sat}}^{2} \left[ A^{\rho_{0}}_{\rho_{0} \rho_{0}}
    + B^{\rho_{0}}_{\rho_{0} \rho_{0}} F_{0} (\xi_{\text{sat}})
    \right]
    + 10 \rho_{\text{sat}}^{2} \xi_{\text{sat}}
    B^{\rho_{0}}_{\rho_{0} \rho_{0}} F_{0}' (\xi_{\text{sat}})
    + \rho_{\text{sat}}^{2} \xi_{\text{sat}}^{2}
    B^{\rho_{0}}_{\rho_{0} \rho_{0}} F_{0}'' (\xi_{\text{sat}})\,,
    \label{eq:K-inf}
\end{align}
where the Fermi momentum at saturation $k_{\text{F,sat}}$ can be obtained from 
$\rho_{\text{sat}}$ through the Thomas-Fermi relation. 

%
%

\subsubsection{Symmetry energy\label{subsubsec:symmetry-energy}}

The isospin symmetry energy $J$ characterises the energy cost to generate a 
finite neutron excess~\cite{BALDO2016203}.
It is defined using the second derivative of the energy per particle
with respect to the isospin asymmetry $I \equiv \rho_{1}/\rho_{0}$
\begin{align}
    S (\rho_{0})
    = \frac{1}{2} \frac{\partial^{2} (E/A)}{\partial I^{2}} \Bigg|_{I=0}
    \leftarrow \rho_{0}^{2} \left[ A^{\rho_{0}}_{\rho_{1} \rho_{1}}
    - B^{\rho_{0}}_{\rho_{0} \rho_{0}} \frac{\xi_{\text{F}}^{2}}{6}\,
    G_{0} (\xi_{\text{F}})
    + B^{\rho_{0}}_{\rho_{1} \rho_{1}} H_{0} (\xi_{\text{F}})
    + B^{\rho_{1}}_{\rho_{0} \rho_{1}} K_{0} (\xi_{\text{F}}) \right]\,,
    \label{eq:symmetry-energy}
\end{align}
from which one gets $J = S(\rho_{\text{sat}})$.
The density-symmetry ($L$) and symmetry-incompressibility 
($K_{\text{sym}}$) coefficients are linked to the slope and 
curvature of the isospin symmetry energy with respect to the density, 
that is~\cite{sadoudi_skyrme_2013,da_costa_impact_2024}
\begin{align}
    L
    = 3 \rho_{0} \frac{\partial S (\rho_{0})}{\partial \rho_{0}}
    \bigg|_{\rho_{0} = \rho_{\text{sat}}}
    & \leftarrow \rho_{\text{sat}}^{2} \Bigg\{
    6 A^{\rho_{0}}_{\rho_{1} \rho_{1}}
    - B^{\rho_{0}}_{\rho_{0} \rho_{0}} \left[
    \frac{4 \xi_{\text{sat}}^{2}}{3}\, G_{0} (\xi_{\text{sat}})
    + \frac{\xi_{\text{sat}}^{3}}{6}\, G_{0}' (\xi_{\text{sat}})
    \right]
    \nonumber \\
    & + B^{\rho_{0}}_{\rho_{1} \rho_{1}} \left[
    6 H_{0} (\xi_{\text{sat}}) + \xi_{\text{sat}} H_{0}' (\xi_{\text{sat}})
    \right] \nonumber \\
    &
    + B^{\rho_{1}}_{\rho_{0} \rho_{1}} \Big[
    6 K_{0} (\xi_{\text{sat}}) + \xi_{\text{sat}} K_{0}' (\xi_{\text{sat}})
    \Big]
    \Bigg\}\,,
    \label{eq:slope-symmetry-energy-L}
\end{align}
and
\begin{align}
    K_{\text{sym}}
    = 9 \rho_{0}^{2} \frac{\partial^{2} S (\rho_{0})}{\partial \rho_{0}^{2}}
    \bigg|_{\rho_{0} = \rho_{\text{sat}}}
    & \leftarrow \rho_{\text{sat}}^{2} \bigg\{
    18 A^{\rho_{0}}_{\rho_{1} \rho_{1}} \nonumber \\
    &\hskip 1cm
    - B^{\rho_{0}}_{\rho_{0} \rho_{0}} \left[
    \frac{20 \xi_{\text{sat}}^{2}}{3}\, G_{0} (\xi_{\text{sat}})
    + \frac{7 \xi_{\text{sat}}^{3}}{3}\, G_{0}' (\xi_{\text{sat}})
    + \frac{\xi_{\text{sat}}^{4}}{6}\, G_{0}'' (\xi_{\text{sat}})
    \right]
    \nonumber \\
    & \hskip 1cm + B^{\rho_{0}}_{\rho_{1} \rho_{1}} \left[
      18 H_{0} (\xi_{\text{sat}}) + 10 \xi_{\text{sat}} H_{0}' (\xi_{\text{sat}})
    + \xi_{\text{sat}}^{2} H_{0}'' (\xi_{\text{sat}})
    \right]
    \nonumber \\
    & \hskip 1cm + B^{\rho_{1}}_{\rho_{0} \rho_{1}} \left[
      18 K_{0} (\xi_{\text{sat}}) + 10 \xi_{\text{sat}} K_{0}' (\xi_{\text{sat}})
    + \xi_{\text{sat}}^2 K_{0}'' (\xi_{\text{sat}})
    \right]
    \bigg\}\,.
    \label{eq:curvature-symmetry-energy-Ksym}
\end{align}
%

%
%

%
\subsection{Decomposition of the EOS in spin-isospin channels\label{subsec:decomposition-eos-ST-channels}}

The projection operators of two particles $i$ and $j$ on spin singlet or spin
triplet states are 
\begin{equation}
    \hat{P}_{ij}^{S=0} = \frac{1}{2} \big(
    \hat{\mathbb{1}}^{\sigma} - \hat{P}_{ij}^{\sigma}
    \big)\,,
    \label{eq:projector-on-S-0}
\end{equation}
and
\begin{equation}
    \hat{P}_{ij}^{S=1} = \frac{1}{2} \big(
    \hat{\mathbb{1}}^{\sigma} + \hat{P}_{ij}^{\sigma}
    \big)\,.
    \label{eq:projector-on-S-1}
\end{equation}
Similar expressions can be written for isospin projection operators.
With these relations, the potential energy can be decomposed in various 
spin-isospin $(S,T)$ channels. 
For a two-body potential, the technique to use in order to perform such 
a decomposition is explained in~\cite{lesinski_isovector_2006}
while analytical expressions up to given order $p$ can be found 
in~\cite{bennaceur_properties_2020}. 

For a three-body interaction, we decide to define the terms of the $(S,T)$ decomposition
as the average of
the potential energies calculated for every possible exchange of
particles, namely
\begin{equation}
    E^{(S,T)} [\rho]
    \equiv \frac{1}{3} \Big\{
    E^{(S,T)}_{12} [\rho] + E^{(S,T)}_{13} [\rho] + E^{(S,T)}_{23} [\rho]
    \Big\}~.
    \label{eq:E-ST-as-mean-value}
\end{equation}
Each of the functionals in the above equation can be calculated as follows
\begin{equation}
    E^{(S,T)}_{ij} [\rho]
    = \frac{1}{6} \int \mathrm{d}x_{1} \mathrm{d}x_{2} \mathrm{d}x_{3}
    \mathrm{d}x_{4} \mathrm{d}x_{5} \mathrm{d}x_{6}~\bar{W}^{(S,T)}_{ij}
    (x_{1}, x_{2}, x_{3} ; x_{4}, x_{5}, x_{6})
    \braket{\Phi | a^{\dagger}_{x_{1}} a^{\dagger}_{x_{2}} a^{\dagger}_{x_{3}}
    a_{x_{6}} a_{x_{5}} a_{x_{4}} | \Phi}\,,
    \label{eq:general-relation-E-ST-3b}
\end{equation}
where $| \Phi\rangle$ is a Hartree-Fock state.
The average $\bar{W}$ is defined as 
\begin{equation}
    \bar{W}^{(S,T)}_{ij} (x_{1}, x_{2}, x_{3} ; x_{4}, x_{5}, x_{6})
    \equiv \frac{1}{6} \sum_{k=1}^{6}
    W_{ij,k}^{(S,T)} (x_{1}, x_{2}, x_{3} ; x_{4}, x_{5}, x_{6})\,,
    \label{eq:definition-average-W-ST-decomposition}
\end{equation}
with
\begin{align}
    & W^{(S,T)}_{ij, 1}
    (x_{1}, x_{2}, x_{3} ; x_{4}, x_{5}, x_{6})
    \equiv \braket{s_{1} q_{1}, s_{2} q_{2}, s_{3} q_{3} |
    \hat{V}_{3\text{b}} (\mathbf{r}_{1}, \mathbf{r}_{2}, \mathbf{r}_{3} ; 
    \mathbf{r}_{4}, \mathbf{r}_{5}, \mathbf{r}_{6})
    \hat{P}^{S}_{ij} \hat{P}^{T}_{ij}
    | s_{4} q_{4}, s_{5} q_{5}, s_{6} q_{6}}\,,
    \label{eq:W1-P-ST} \\
    & W^{(S,T)}_{ij, 2}
    (x_{1}, x_{2}, x_{3} ; x_{4}, x_{5}, x_{6})
    \equiv \braket{s_{1} q_{1}, s_{3} q_{3}, s_{2} q_{2} |
    \hat{V}_{3\text{b}} (\mathbf{r}_{1}, \mathbf{r}_{3}, \mathbf{r}_{2} ; 
    \mathbf{r}_{4}, \mathbf{r}_{6}, \mathbf{r}_{5})
    \hat{P}^{S}_{ij} \hat{P}^{T}_{ij}
    | s_{4} q_{4}, s_{6} q_{6}, s_{5} q_{5}}\,,
    \label{eq:W2-P-ST} \\
    & W^{(S,T)}_{ij, 3}
    (x_{1}, x_{2}, x_{3} ; x_{4}, x_{5}, x_{6})
    \equiv \braket{s_{2} q_{2}, s_{1} q_{1}, s_{3} q_{3} |
    \hat{V}_{3\text{b}} (\mathbf{r}_{2}, \mathbf{r}_{1}, \mathbf{r}_{3} ; 
    \mathbf{r}_{5}, \mathbf{r}_{4}, \mathbf{r}_{6})
    \hat{P}^{S}_{ij} \hat{P}^{T}_{ij}
    | s_{5} q_{5}, s_{4} q_{4}, s_{6} q_{6}}\,,
    \label{eq:W3-P-ST} \\
    & W^{(S,T)}_{ij, 4}
    (x_{1}, x_{2}, x_{3} ; x_{4}, x_{5}, x_{6})
    \equiv \braket{s_{2} q_{2}, s_{3} q_{3}, s_{1} q_{1} |
    \hat{V}_{3\text{b}} (\mathbf{r}_{2}, \mathbf{r}_{3}, \mathbf{r}_{1} ; 
    \mathbf{r}_{5}, \mathbf{r}_{6}, \mathbf{r}_{4})
    \hat{P}^{S}_{ij} \hat{P}^{T}_{ij}
    | s_{5} q_{5}, s_{6} q_{6}, s_{4} q_{4}}\,,
    \label{eq:W4-P-ST} \\
    & W^{(S,T)}_{ij, 5}
    (x_{1}, x_{2}, x_{3} ; x_{4}, x_{5}, x_{6})
    \equiv \braket{s_{3} q_{3}, s_{1} q_{1}, s_{2} q_{2} |
    \hat{V}_{3\text{b}} (\mathbf{r}_{3}, \mathbf{r}_{1}, \mathbf{r}_{2} ; 
    \mathbf{r}_{6}, \mathbf{r}_{4}, \mathbf{r}_{5})
    \hat{P}^{S}_{ij} \hat{P}^{T}_{ij}
    | s_{6} q_{6}, s_{4} q_{4}, s_{5} q_{5}}\,,
    \label{eq:W5-P-ST} \\
    & W^{(S,T)}_{ij, 6}
    (x_{1}, x_{2}, x_{3} ; x_{4}, x_{5}, x_{6})
    \equiv \braket{s_{3} q_{3}, s_{2} q_{2}, s_{1} q_{1} |
    \hat{V}_{3\text{b}} (\mathbf{r}_{3}, \mathbf{r}_{2}, \mathbf{r}_{1} ; 
    \mathbf{r}_{6}, \mathbf{r}_{5}, \mathbf{r}_{4})
    \hat{P}^{S}_{ij} \hat{P}^{T}_{ij}
    | s_{6} q_{6}, s_{5} q_{5}, s_{4} q_{4}}\,,
    \label{eq:W6-P-ST}
\end{align}
for $\hat{V}_{3\text{b}}$ given by~\eqref{eq:final-form-V-3b}.
Applying the prescription~\eqref{eq:E-ST-as-mean-value} for symmetric 
nuclear matter yields the $(S,T)$ decomposition of the potential energy in 
the four channels
\begin{align}
    \frac{E^{(0,0)}}{A}
    &\leftarrow \frac{2 W_{3,1} - W_{3,2}}{384}\, \rho_{0}^{2} \big[
    1 - F_{0} (\xi_{\text{F}})
    \big]\,,
    \label{eq:eos-3b-00-SNM} \\
    \frac{E^{(0,1)}}{A}
    &\leftarrow \frac{2 W_{3,1} - W_{3,2}}{128}\, \rho_{0}^{2} \big[
    1 + F_{0} (\xi_{\text{F}})
    \big]\,,
    \label{eq:eos-3b-01-SNM} \\
\frac{E^{(1,0)}}{A}
    &\leftarrow \frac{1}{384}\, \rho_{0}^{2} \Big\{
    2 W_{3,1} \big[
    11 - 5 F_{0} (\xi_{\text{F}})
    \big]
    + W_{3,2} \big[
    13 - 19 F_{0} (\xi_{\text{F}})
    \big]
    \Big\}\,,
    \label{eq:eos-3b-10-SNM} \\
    \frac{E^{(1,1)}}{A}
    &\leftarrow \frac{6W_{3,1} + 5 W_{3,2}}{128}\, \rho_{0}^{2} \big[
    1 - F_{0} (\xi_{\text{F}})
    \big]\,.
    \label{eq:eos-3b-11-SNM}
\end{align}

\subsection{Landau parameters\label{subsec:landau-parameters}}

Landau parameters allow for a simple parametrization of the residual interaction in infinite homogeneous nuclear matter~\cite{TOWNER1987263}.
The basic ingredient needed to compute Landau parameters from the central three-body
interaction considered in this work is the residual
particle-hole interaction of INM~\cite{sadoudi_skyrme_2013}. 
The latter is obtained by calculating the second order functional 
derivative of the ph EDF
\begin{align}
    V_{\text{ph, res.}} (x_{1}, x_{2} ; x_{3}, x_{4})
    = \braket{x_{1}, x_{2} | \hat{V}_{\text{ph, res.}} | x_{3}, x_{4}}
    = \frac{\delta^{2} E_{\text{ph}}}{
    \delta \rho (x_{3}, x_{1}) \delta \rho (x_{4}, x_{2})}\,,
    \label{eq:definition-residual-interaction}
\end{align}
with conventional notation $x_{i} = (\mathbf{r}_{i} s_{i} q_{i})$ for 
generalised coordinates. We can recall that
even when $E_{\text{ph}}$ is obtained from a three-body pseudopotential, 
the residual interaction defined by~(\ref{eq:definition-residual-interaction})
remains of course a two-body force.

It is possible to decompose the above interaction in spin-isospin channels $(S,T)$
whose matrix elements in momentum space are then written
as~\cite{idini_landau_2017}
\begin{align}
    V_{\text{ph, res.}}^{(S,T)}
    (\mathbf{k}_{1}, \mathbf{k}_{2} ; \mathbf{k}_{3}, \mathbf{k}_{4})
    = \braket{\mathbf{k}_{1}, \mathbf{k}_{2} | \hat{V}_{\text{ph, res.}}^{(S,T)} |
    \mathbf{k}_{3}, \mathbf{k}_{4}}
    = \frac{\delta (\mathbf{q} - \mathbf{q}^{\prime})}{(2\pi)^{3}}
    \,\mathcal{V}_{\text{ph, res.}}^{(S,T)}
    (q, \mathbf{k} - \mathbf{k}^{\prime})\,,
    \label{eq:definition-fourier-transform-residual-interaction}
\end{align}
where $\mathbf{k}_{1} = \mathbf{k} + \mathbf{q},~\mathbf{k}_{2}
= \mathbf{k}^{\prime},~\mathbf{k}_{3} = \mathbf{k}$ and 
$\mathbf{k}_{4} = \mathbf{k}^{\prime} + \mathbf{q}^{\prime}$ with notations
from~\cite{idini_landau_2017} for the transferred momentum $\mathbf q$.
Finally, the Landau parameters $f_{\ell}^{(S,T)}$ are given as the
coefficients of the residual interaction expanded over Legendre polynomials
in different $(S,T)$ channels for $q=\left\Vert\mathbf q\right\Vert
\to0$ (see {\em e.g.}~\cite{lesinski_isovector_2006,davesne_linear_2021,idini_landau_2017}),
namely
\begin{align}
    \lim\limits_{q \to 0}~\mathcal{V}_{\text{ph, res.}}^{(S,T)}
    (q, \mathbf{k} - \mathbf{k}^{\prime})
    \Big. \Big|_{|| \mathbf{k} || = || \mathbf{k}^{\prime} || = k_{\text{F}}}
    = \sum_{\ell=0}^{\infty}\, f_{\ell}^{(S,T)} (k_{\text{F}})
    P_{\ell} \big( \hat{k} \cdot \hat{k}^{\prime} \big)\,.
    \label{eq:expansion-legendre-polynomials-residual-interaction}
\end{align}
The Landau parameters for the two-body part of the pseudopotential discussed 
in this work can be found in~\cite{idini_landau_2017}.
For the contributions from the three-body part, calculations are very similar to that of
the LO two-body term and these contributions are
\begin{align}
    f_{\ell}^{(0,0)} (k_{\text{F}})
    &\leftarrow 6 A^{\rho_{0}}_{\rho_{0} \rho_{0}} \rho_{0} \delta_{\ell 0}
    + 4 B^{\rho_{0}}_{\rho_{0} \rho_{0}} \rho_{0}
    K_{0} (\xi_{\text{F}}) \delta_{\ell 0}
    + 2 B^{\rho_{0}}_{\rho_{0} \rho_{0}} \rho_{0}
    \mathrm{e}^{-\frac{\xi_{\text{F}}^{2}}{2}}
    (2\ell+1) \,i_{\ell\!}\left(\tfrac{\xi_{\text{F}}^{2}}{2}\right)\,,
    \label{eq:landau-parameter-ST-00} \\
    f_{\ell}^{(1,0)} (k_{\text{F}})
    &\leftarrow 2 A^{\rho_{0}}_{\mathbf{s}_{0} \mathbf{s}_{0}} \rho_{0} \delta_{\ell 0}
    + A^{\mathbf{s}_{0}}_{\rho_{0} \mathbf{s}_{0}} \rho_{0} \delta_{\ell 0}
    + 2 B^{\mathbf{s}_{0}}_{\rho_{0} \mathbf{s}_{0}} \rho_{0}
    K_{0} (\xi_{\text{F}}) \delta_{\ell 0}
    + 2 B^{\rho_{0}}_{\mathbf{s}_{0} \mathbf{s}_{0}} \rho_{0}
    \mathrm{e}^{-\frac{\xi_{\text{F}}^{2}}{2}}
    (2\ell+1) \,i_{\ell\!} \left(\tfrac{\xi_{\text{F}}^{2}}{2}\right)\,,
    \label{eq:landau-parameter-ST-10} \\
    f_{\ell}^{(0,1)} (k_{\text{F}})
    &\leftarrow 2 A^{\rho_{0}}_{\rho_{1} \rho_{1}} \rho_{0} \delta_{\ell 0}
    + 2 B^{\rho_{1}}_{\rho_{0} \rho_{1}} \rho_{0}
    K_{0} (\xi_{\text{F}}) \delta_{\ell 0}
    + 2 B^{\rho_{0}}_{\rho_{1} \rho_{1}} \rho_{0}
    \mathrm{e}^{-\frac{\xi_{\text{F}}^{2}}{2}}
    (2\ell+1) \,i_{\ell\!} \left(\tfrac{\xi_{\text{F}}^{2}}{2}\right)\,,
    \label{eq:landau-parameter-ST-01} \\
    f_{\ell}^{(1,1)} (k_{\text{F}})
    &\leftarrow 2 A^{\rho_{0}}_{\mathbf{s}_{1} \mathbf{s}_{1}} \rho_{0} \delta_{\ell 0}
    + 2 B^{\mathbf{s}_{1}}_{\rho_{0} \mathbf{s}_{1}} \rho_{0}
    K_{0} (\xi_{\text{F}}) \delta_{\ell 0}
    + 2 B^{\rho_{0}}_{\mathbf{s}_{1} \mathbf{s}_{1}} \rho_{0}
    \mathrm{e}^{-\frac{\xi_{\text{F}}^{2}}{2}}
    (2\ell+1) \,i_{\ell\!} \left(\tfrac{\xi_{\text{F}}^{2}}{2}\right)\,,
    \label{eq:landau-parameter-ST-11}
\end{align}
where $i_{\ell}$ is the modified spherical Bessel function of the first 
kind.


\section{Expressions for a spherically-symmetric system\label{sec:spherical-symmetry}}

\subsection{Wave functions and densities\label{subsec:introduction-spherical-symmetry}}

In this section, we assume that the solutions of the HFB equations are
invariant under rotations and parity inversion in coordinate space and 
that they do not mix protons and neutrons.
The eigenvectors of the HFB matrix can then be labeled with the quantum numbers
$(n,\ell,j,m,q)$.
Under this assumption, the two components $U$ and $V$ of the HFB spinors
read~\cite{schunck_energy_2019,decharge_hartree-fock-bogolyubov_1980,dobaczewski_hartree-fock-bogolyubov_1984,bennaceur_coordinate-space_2005}
\begin{align}
    U_{n \ell jm} (\mathbf{r} m_{s} q) &= \frac{1}{r} \sum_{m_{\ell}}
    \braket{\ell m_{\ell}, s m_{s} | jm}\,
    u_{q n \ell j} (r)\, Y^{(\ell)}_{m_{\ell}} (\hat{r})\,,
    \label{eq:component-U-HFB-spinors-spherical-symmetry} \\
    V_{n \ell jm} (\mathbf{r} m_{s} q) &= \frac{1}{r} \sum_{m_{\ell}}
    \braket{\ell m_{\ell}, s m_{s} | jm}\,
    v_{q n \ell j} (r)\, Y^{(\ell)}_{m_{\ell}} (\hat{r})\,,
    \label{eq:component-V-HFB-spinors-spherical-symmetry}
\end{align}
where $s = \frac{1}{2}$, $m_{s} = \pm s$ and $m_{\ell} \in \{ - \ell, \ldots, +\ell \}$. 

From these two expressions, one can express the various densities entering
the EDF kernel.
For instance, the non-local density reads 
\begin{align}
    \rho_{q} (\mathbf{r}_{1}, \mathbf{r}_{2})
    = \sum_{n \ell jm m_{s}} V^{*}_{n \ell jm} (\mathbf{r}_{2} m_{s} q)
    V_{n \ell jm} (\mathbf{r}_{1} m_{s} q)
    = \sum_{\ell j m_{\ell}} \rho_{q\ell j} (r_{1}, r_{2})
    \,Y^{(\ell)}_{m_{\ell}} (\hat{r}_{1})\, {Y^{(\ell)}_{m_{\ell}}}^*\!\! (\hat{r}_{2})\,,
    \label{eq:multipole-expansion-non-local-normal-density-r1-r2}
\end{align}
where we define the multipoles of the scalar density as
\begin{equation}
    \rho_{q\ell j} (r_{1}, r_{2})
    = \rho_{q\ell j} (r_{2}, r_{1})
    = \frac{2j+1}{2\ell+1} \sum_{n}
    \frac{v_{qn\ell j} (r_{1}) v_{qn\ell j} (r_{2})}{r_{1}r_{2}} \,,
    \label{eq:reduced-non-local-normal-density-r1-r2}
\end{equation}
with a local part that can then be written as 
\begin{align}
    \rho_{q} (r)
    = \sum_{\ell j} \frac{2\ell+1}{4\pi} \rho_{q \ell j} (r)\,.
    \label{eq:multipole-expansion-local-normal-density}
\end{align}
%
%
The non-local pairing density is given by
\begin{align}
    \tilde{\rho}_{q} (\mathbf{r}_{1}, \mathbf{r}_{2})
    = -\sum_{n\ell jm m_{s}} V^{*}_{n\ell jm} (\mathbf{r}_{2} m_{s} q)
    U_{n\ell jm} (\mathbf{r}_{1} m_{s} q)
    = \sum_{\ell j m_{\ell}} \tilde{\rho}_{q\ell j} (r_{1}, r_{2})
    \,Y^{(\ell)}_{m_{\ell}} (\hat{r}_{1})\,{Y^{(\ell)}_{m_{\ell}}}^*\!\!(\hat{r}_{2})\,,
    \label{eq:multipole-expansion-non-local-pair-density-r1-r2}
\end{align}
with
\begin{equation}
    \tilde{\rho}_{q\ell j} (r_{1}, r_{2})
    = -\frac{2j+1}{2\ell+1}  \sum_{n}
    \frac{u_{qn\ell j} (r_{1}) v_{qn\ell j} (r_{2})}{r_{1}r_{2}}\,.
    \label{eq:reduced-non-local-pair-density-r1-r2}
\end{equation}
The expression of the local pairing density is not needed because the
three-body pseudopotential has been developed in such a way that
$\tilde{\rho}_{q} (\mathbf{r})$ does not appear in the EDF.


The other densities entering the energy are the normal and
pairing spin-density that both can be written as spherical rank-1 tensor
densities whose components are
\begin{align}
     s^{(1)}_{q,\mu} (\mathbf{r}_{1}, \mathbf{r}_{2})
    &= \sum_{\substack{n\ell jm \\ m_{s_1} m_{s_2}}}
    V_{n\ell jm}^{*} (\mathbf{r}_{2} m_{s_{2}} q)
    V_{n\ell jm} (\mathbf{r}_{1} m_{s_{1}} q)
    \braket{s m_{s_{2}} | \sigma_{\mu}^{(1)} | s m_{s_{1}}}\,,
    \nonumber \\
    & = \sum_{\ell j} \sqrt{\frac{2\ell+1}{\ell(\ell+1)}} f(j,\ell,s)
    \rho_{q\ell j} (r_{1}, r_{2}) \nonumber \\
    &\hskip 2cm\times\sum_{m_{\ell_{1}} m_{\ell_{2}}}
    (-1)^{\ell+m_{\ell_{1}}} 
    \begin{pmatrix}
        \ell & \ell & 1 \\
        m_{\ell_{2}} & -m_{\ell_{1}} & \mu
    \end{pmatrix}
    Y^{(\ell)}_{m_{\ell_{1}}} (\hat{r}_{1})
    Y^{(\ell)*}_{m_{\ell_{2}}} (\hat{r}_{2})\,,
    \label{eq:vector-isoscalar-density-spherical-symmetry-r1-r2}
\end{align}
for the normal spin-density and
\begin{align}
     \tilde{s}^{(1)}_{q,\mu} (\mathbf{r}_{1}, \mathbf{r}_{2})
    &= - \sum_{\substack{n\ell jm \\ m_{s_1} m_{s_2}}}
    V^{*}_{n\ell jm} (\mathbf{r}_{2} m_{s_{2}} q)
    U_{n\ell jm} (\mathbf{r}_{1} m_{s_{1}} q)
    \braket{s m_{s_{2}} | \sigma_{\mu}^{(1)} | s m_{s_{1}}}\,,
    \nonumber \\
    & = \sum_{\ell j} \sqrt{\frac{2\ell+1}{\ell(\ell+1)}} f(j,\ell,s)
    \tilde{\rho}_{q\ell j} (r_{1}, r_{2}) \nonumber \\
    &\hskip 2cm\times\sum_{m_{\ell_{1}} m_{\ell_{2}}}
    (-1)^{\ell+m_{\ell_{1}}} 
    \begin{pmatrix}
        \ell & \ell & 1 \\
        m_{\ell_{2}} & -m_{\ell_{1}} & \mu
    \end{pmatrix}
    Y^{(\ell)}_{m_{\ell_{1}}} (\hat{r}_{1})
    Y^{(\ell)*}_{m_{\ell_{2}}} (\hat{r}_{2})\,.
    \label{eq:pair-spin-density-spherical-symmetry-r1-r2}
\end{align}
for the pairing spin-density.

For convenience in the following, we also introduced the function $f$ defined as 
\begin{align}
    f (x, y, z)
    = x (x + 1) - y (y + 1) - z (z + 1)~.
    \label{eq:definition-f-xyz}
\end{align}
Note that, for a stationary spherical state, all time-odd densities are 
zero which implies $\mathbf{s}_{q} (\mathbf{r}) = \mathbf{0}$.
Additionally, one always has $\tilde{\mathbf{s}}_{q} (\mathbf{r})
= \mathbf{0}$ as a trivial consequence of the Pauli principle.

%
%

For sake of completeness, we also recall the multipole expansion of the 
Gaussian form factor entering the EDF~\cite{chiu_translational_1994}
\begin{align}
    g_{a} (\mathbf{r}_{12})
    = \sum_{\ell m_{\ell}}
    g_{a}^{(\ell)} (r_{12}) Y^{(\ell)*}_{m_{\ell}} (\hat{r}_{1})
    Y^{(\ell)}_{m_{\ell}} (\hat{r}_{2})\,,
    \label{eq:multipole-expansion-gaussian}
\end{align}
with the multipoles of $g_a$ given by
\begin{equation}
    g_{a}^{(\ell)} (r_{12})
    = \frac{
    4 \,\mathrm{e}^{-\frac{\mathbf{r}_{1}^{2} + \mathbf{r}_{2}^{2}}{a^{2}}}
    }{
    a^{3} \sqrt{\pi}
    } \,
    i_{\ell} \left( \frac{2 r_{1} r_{2}}{a^{2}} \right)\,,
    \label{eq:definition-g-a-l}
\end{equation}
where $i_\ell$ is a modified spherical Bessel function of the first kind.

\subsection{Energy density functional\label{subsec:edf-spherical-symmetry}}

In this section we provide the expression for the energy in the ph (particle-hole) and pp
(particle-particle)
channels after integration over the angles has been performed. 
This integration can be done using standard techniques of angular momentum 
theory~\cite{brussaard_shell-model_1977,khersonskii_quantum_1989}. 
Note that the summation over $\ell$ and $j$ appearing in
equation~\eqref{eq:multipole-expansion-local-normal-density} does not
need to be written.
This means that we do not explicitly write the multipole expansion of the
local density and, since it has no angular dependence, we note
$\rho_{q} (\mathbf{r}) \equiv \rho_{q} (r)$. 

%
%

\subsubsection{Particle-hole channel\label{subsubsec:ph-channel-edf-spherical-symmetry}}

The local term for a nucleus with spherical symmetry reads 
\begin{align}
    E_{\text{ph}}^{\text{loc}}
    = 4\pi \int\! \mathrm{d}r_{1}\, \mathrm{d}r_{2}\,r_{1}^{2} r_{2}^{2}
    \,g_{a}^{(0)} (r_{1}, r_{2}) \rho_{0} (r_{1}) \left[
    \left( A^{\rho_{0}}_{\rho_{0}\rho_{0}}
    - A^{\rho_{0}}_{\rho_{1}\rho_{1}} \right)
    \rho_{0}^{2} (r_{2})
    + 2 A^{\rho_{0}}_{\rho_{1}\rho_{1}}
    \sum_{q} \rho_{q}^{2} (r_{2})
    \right]\,.
    \label{eq:local-term-edf-spherical-symmetry}
\end{align}

%
%

The parts of the EDF involving the local scalar-isoscalar and
scalar-isovector densities multiplied by non-local densities are
respectively written 
\begin{align}
    E_{\text{ph}}^{\rho_{0}}
    & = \frac{1}{4\pi} \sum_{\substack{L \ell_{1} j_{1}\\ \ell_{2} j_{2}}}
    (2L+1) (2\ell_{1}+1) (2\ell_{2}+1)
    \begin{pmatrix}
        \ell_{1} & \ell_{2} & L \\
        0 & 0 & 0
    \end{pmatrix}^{2}
    \int\! \mathrm{d}r_{1}\, \mathrm{d}r_{2}\,r_{1}^{2} r_{2}^{2}
    \,g_{a}^{(L)} (r_{1}, r_{2}) \rho_{0} (r_{1})
    \nonumber \\
    & \times \left[
    \left( B^{\rho_{0}}_{\rho_{0}\rho_{0}}
    - B^{\rho_{0}}_{\rho_{1}\rho_{1}} \right)
    \rho_{0 \ell_{1} j_{1}} (r_{1}, r_{2})
    \rho_{0 \ell_{2} j_{2}} (r_{2}, r_{1})
    + 2 B^{\rho_{0}}_{\rho_{1}\rho_{1}}
    \sum_{q} \rho_{q \ell_{1} j_{1}} (r_{1}, r_{2})
    \rho_{q \ell_{2} j_{2}} (r_{2}, r_{1})
    \right]\,,
    \label{eq:non-loc-rho0-term-edf-spherical-symmetry}
\end{align}
and
\begin{align}
    E_{\text{ph}}^{\rho_{1}}
    &= \frac{1}{4\pi} \sum_{\substack{L \ell_{1} j_{1}\\ \ell_{2} j_{2}}}
    (2L+1) (2\ell_{1}+1) (2\ell_{2}+1)
    \begin{pmatrix}
        \ell_{1} & \ell_{2} & L \\
        0 & 0 & 0
    \end{pmatrix}^{2}
    \nonumber \\
    & \times \int\! \mathrm{d}r_{1}\, \mathrm{d}r_{2}\,r_{1}^{2} r_{2}^{2}\,
    g_{a}^{(L)} (r_{1}, r_{2}) \rho_{1} (r_{1})
    B^{\rho_{1}}_{\rho_{0} \rho_{1}} \left[
    \rho_{\text{n} \ell_{1} j_{1}} (r_{1}, r_{2})
    \rho_{\text{n} \ell_{2} j_{2}} (r_{2}, r_{1})
    - \rho_{\text{p} \ell_{1} j_{1}} (r_{1}, r_{2})
    \rho_{\text{p} \ell_{2} j_{2}} (r_{2}, r_{1})
    \right]\,.
    \label{eq:non-loc-rho1-term-edf-spherical-symmetry}
\end{align}
%

%
%

Finally, the parts of the EDF involving the local scalar-isoscalar and
scalar-isovector densities multiplied by non-local spin densities are
\begin{align}
    E^{\mathbf{s}_{0}}_{\text{ph}}
    & = \frac{1}{8\pi}
    \sum_{\substack{L \ell_{1} j_{1}\\ \ell_{2} j_{2}}}
    \frac{(2L+1) (2\ell_{1}+1) (2\ell_{2}+1)}{
    \ell_{1} (\ell_{1}+1) \ell_{2} (\ell_{2}+1)}
    f(j_{1},\ell_{1},s) f(j_{2},\ell_{2},s) f(\ell_{1},\ell_{2},L)
    \begin{pmatrix}
        \ell_{1} & \ell_{2} & L \\
        0 & 0 & 0
    \end{pmatrix}^{2}
    \nonumber \\
    & \times \int\! \mathrm{d}r_{1}\, \mathrm{d}r_{2}\,r_{1}^{2} r_{2}^{2}\,
    g_{a}^{(L)} (r_{1}, r_{2}) \rho_{0} (r_{1}) \Biggl[
    \left( B^{\rho_{0}}_{\mathbf{s}_{0}\mathbf{s}_{0}}
    - B^{\rho_{0}}_{\mathbf{s}_{1}\mathbf{s}_{1}} \right)
    \rho_{0 \ell_{1} j_{1}} (r_{1}, r_{2})
    \rho_{0 \ell_{2} j_{2}} (r_{2}, r_{1})
    \nonumber \\
    & \hspace{3cm}+ 2 B^{\rho_{0}}_{\mathbf{s}_{1}\mathbf{s}_{1}}
    \sum_{q} \rho_{q \ell_{1} j_{1}} (r_{1}, r_{2})
    \rho_{q \ell_{2} j_{2}} (r_{2}, r_{1})\Biggr]\,,
    \label{eq:non-loc-s0-term-edf-spherical-symmetry}
\end{align}
and
\begin{align}
     E^{\mathbf{s}_{1}}_{\text{ph}}
    &= \frac{1}{16\pi}
    \sum_{\substack{L \ell_{1} j_{1}\\ \ell_{2} j_{2}}}
    \frac{(2L+1) (2\ell_{1}+1) (2\ell_{2}+1)}{
    \ell_{1} (\ell_{1}+1) \ell_{2} (\ell_{2}+1)}
    f(j_{1},\ell_{1},s) f(j_{2},\ell_{2},s) f(\ell_{1},\ell_{2},L)
    \begin{pmatrix}
        \ell_{1} & \ell_{2} & L \\
        0 & 0 & 0
    \end{pmatrix}^{2}
    \nonumber \\
    & \times \int\! \mathrm{d}r_{1}\, \mathrm{d}r_{2}\,r_{1}^{2} r_{2}^{2}\,
    g_{a}^{(L)} (r_{1}, r_{2}) \rho_{1} (r_{1})
    B^{\rho_{1}}_{\mathbf{s}_{0}\mathbf{s}_{1}} \Big[
    \rho_{0 \ell_{1} j_{1}} (r_{1}, r_{2})
    \rho_{1 \ell_{2} j_{2}} (r_{2}, r_{1})
    + \rho_{1 \ell_{1} j_{1}} (r_{1}, r_{2})
    \rho_{0 \ell_{2} j_{2}} (r_{2}, r_{1})
    \Big]\,.
    \label{eq:non-loc-s1-term-edf-spherical-symmetry}
\end{align}
%

%
%

\subsubsection{Particle-particle channel\label{subsubsec:pp-channel-edf-spherical-symmetry}}

For a spherical nucleus, the contributions to the energy from the terms with
scalar and vector pairing densities are
\begin{align}
    E^{\tilde{\rho}_{q}}_{\text{pp}} 
    & = \frac{1}{4\pi} 
    \sum_{L \ell_{1} j_{1} \ell_{2} j_{2}}
    (2L+1) (2\ell_{1}+1) (2\ell_{2}+1)
    \begin{pmatrix}
        \ell_{1} & \ell_{2} & L \\
        0 & 0 & 0
    \end{pmatrix}^{2}
    \nonumber \\
    &\times \int\! \mathrm{d}r_{1}\, \mathrm{d}r_{2}\,r_{1}^{2} r_{2}^{2}\,
    g_{a}^{(L)} (r_{1}, r_{2}) C^{\rho}_{\tilde{\rho}\tilde{\rho}}
    \sum_{q} \frac{\rho_{\bar{q}} (r_{1}) + \rho_{\bar{q}} (r_{2})}{2}\,
    \tilde{\rho}_{q\ell_{1}j_{1}} (r_{1}, r_{2})
    \tilde{\rho}_{q\ell_{2}j_{2}} (r_{2}, r_{1})\,,
    \label{eq:non-loc-rhotq-term-edf-spherical-symmetry}
\end{align}
and
\begin{align}
    E^{\tilde{\mathbf{s}}_{q}}_{\text{pp}} 
    & = \frac{1}{8\pi}
    \sum_{L \ell_{1} j_{1} \ell_{2} j_{2}}
    \frac{(2L+1) (2\ell_{1}+1) (2\ell_{2}+1)}{
    \ell_{1} (\ell_{1}+1) \ell_{2} (\ell_{2}+1)}
    f(j_{1},\ell_{1},s) f(j_{2},\ell_{2},s) f(\ell_{1},\ell_{2},L)
    \begin{pmatrix}
        \ell_{1} & \ell_{2} & L \\
        0 & 0 & 0
    \end{pmatrix}^{2}
    \nonumber \\
    &
    \times \int\! \mathrm{d}r_{1}\, \mathrm{d}r_{2}\,r_{1}^{2} r_{2}^{2}\,
    g_{a}^{(L)} (r_{1}, r_{2}) C^{\rho}_{\tilde{\mathbf{s}}\tilde{\mathbf{s}}}
    \sum_{q} \frac{\rho_{\bar{q}} (r_{1}) + \rho_{\bar{q}} (r_{2})}{2}\,
    \tilde{\rho}_{q\ell_{1}j_{1}} (r_{1}, r_{2})
    \tilde{\rho}_{q\ell_{2}j_{2}} (r_{2}, r_{1})\,.
    \label{eq:non-loc-stq-term-edf-spherical-symmetry}
\end{align}
%

%
%

%
\subsection{Mean fields\label{subsec:mean-fields-spherical-symmetry}}

The mean field and pairing mean field are obtained from the variation of the 
EDF with respect to the reduced
densities~\eqref{eq:reduced-non-local-normal-density-r1-r2}
and~\eqref{eq:reduced-non-local-pair-density-r1-r2}, that is 
\begin{align}
    h_{q \ell j} (r_{1}, r_{2})
    = \frac{\delta E}{\delta \rho_{q \ell j} (r_2, r_1)}\,,
    \label{eq:definition-mean-field-spherical-symmetry}
\end{align}
and
\begin{align}
    \tilde{h}_{q \ell j} (r_{1}, r_{2})
    = \frac{\delta E}{\delta \tilde{\rho}_{q \ell j} (r_2, r_1)}\,.
    \label{eq:definition-pairing-mean-field-spherical-symmetry}
\end{align}
%

%
%

\subsubsection{Mean field\label{subsubsec:mean-field-spherical-symmetry}}

The variation of~\eqref{eq:local-term-edf-spherical-symmetry} yields
\begin{align}
    h_{q \ell j}^{\text{loc}} (r_{1})
    & = \int\! \mathrm{d}r_{2}\,r_{2}^{2}\, g_{a}^{(0)} (r_{1}, r_{2})
    \Biggl\{ \big( A^{\rho_{0}}_{\rho_{0}\rho_{0}}
    - A^{\rho_{0}}_{\rho_{1}\rho_{1}} \big) \left[
    \rho_{0}^{2} (r_{2})
    + 2 \rho_{0} (r_{1}) \rho_{0} (r_{2})
    \right]
    \nonumber \\
    & \hspace{2cm}
    + 2 A^{\rho_{0}}_{\rho_{1}\rho_{1}} 
    \biggl[2 \rho_{0} (r_{2}) \rho_{q} (r_{1})
    +  \sum_{q^{\prime}} \rho_{q^{\prime}}^{2} (r_{2})
    \biggr]
    \Biggr\}\,.
    \label{eq:local-mean-field-spherical-symmetry}
\end{align}
%
%
The variation of~\eqref{eq:non-loc-rho0-term-edf-spherical-symmetry} splits 
into a local contribution
\begin{align}
    h_{q \ell j}^{\rho_{0},\text{loc}} (r_{1})
    & = \frac{1}{16\pi^{2}}
    \sum_{\substack{L \ell_{1} j_{1}\\ \ell_{2} j_{2}}}
    (2L+1) (2\ell_{1}+1) (2\ell_{2}+1)
    \begin{pmatrix}
        \ell_{1} & \ell_{2} & L \\
        0 & 0 & 0
    \end{pmatrix}^{2}
    \int\! \mathrm{d}r_{2}\,r_{2}^{2}\, g_{a}^{(L)} (r_{1}, r_{2})
    \nonumber \\
    & \times \left[
    \left( B^{\rho_{0}}_{\rho_{0}\rho_{0}}
    - B^{\rho_{0}}_{\rho_{1}\rho_{1}} \right)
    \rho_{0 \ell_{1} j_{1}} (r_{1}, r_{2})
    \rho_{0 \ell_{2} j_{2}} (r_{2}, r_{1})
    + 2 B^{\rho_{0}}_{\rho_{1}\rho_{1}} \sum_{q^{\prime}}
    \rho_{q^{\prime} \ell_{1} j_{1}} (r_{1}, r_{2})
    \rho_{q^{\prime} \ell_{2} j_{2}} (r_{2}, r_{1})
    \right]\,,
    \label{eq:local-part-mean-field-rho0-spherical-symmetry}
\end{align}
and a non-local one 
\begin{align}
    h_{q \ell j}^{\rho_{0},\text{nloc}} (r_{1}, r_{2})
    & = \frac{1}{4\pi} \sum_{L \ell_{2} j_{2}}
    (2L+1) (2\ell_{2}+1)
    \begin{pmatrix}
        \ell & \ell_{2} & L \\
        0 & 0 & 0
    \end{pmatrix}^{2}
    r_{1} r_{2}\, g_{a}^{(L)} (r_{1}, r_{2})
    \,\frac{\rho_{0} (r_{1}) + \rho_{0} (r_{2})}{2}
    \nonumber \\
    & \times \left[
    2 \big( B^{\rho_{0}}_{\rho_{0}\rho_{0}}
    - B^{\rho_{0}}_{\rho_{1}\rho_{1}} \big)
    \rho_{0 \ell_{2} j_{2}} (r_{1}, r_{2})
    + 4 B^{\rho_{0}}_{\rho_{1}\rho_{1}}
    \rho_{q \ell_{2} j_{2}} (r_{1}, r_{2})
    \right]\,.
    \label{eq:non-local-part-mean-field-rho0-spherical-symmetry}
\end{align}

%
%

The variation of~\eqref{eq:non-loc-rho1-term-edf-spherical-symmetry} also 
yields a local contribution
\begin{align}
    h^{\rho_{1},\text{loc}}_{q \ell j} (r_{1})
    & = \frac{1}{16\pi^{2}}
    \sum_{\substack{L \ell_{1} j_{1}\\ \ell_{2} j_{2}}}
    (2L+1) (2\ell_{1}+1) (2\ell_{2}+1)
    \begin{pmatrix}
        \ell_{1} & \ell_{2} & L \\
        0 & 0 & 0
    \end{pmatrix}^{2}
    \int\! \mathrm{d}r_{2}\,r_{2}^{2}\,
    g_{a}^{(L)} (r_{1}, r_{2}) B^{\rho_{1}}_{\rho_{0} \rho_{1}}
    \nonumber \\
    & \times \left[ 4 \rho_{q} (r_{1}) - 2 \rho_{0} (r_{1})
    \right] \left[
    \rho_{\text{n} \ell_{1} j_{1}} (r_{1}, r_{2})
    \rho_{\text{n} \ell_{2} j_{2}} (r_{2}, r_{1})
    - \rho_{\text{p} \ell_{1} j_{1}} (r_{1}, r_{2})
    \rho_{\text{p} \ell_{2} j_{2}} (r_{2}, r_{1})
    \right]\,,
    \label{eq:local-part-mean-field-rho1-spherical-symmetry}
\end{align}
and a non-local one 
\begin{align}
    h^{\rho_{1},\text{nloc}}_{q \ell j} (r_{1}, r_{2})
    & = -\frac{2}{\pi} \sum_{L \ell_{2} j_{2}}
    (2L+1) (2\ell_{2}+1)
    \begin{pmatrix}
        \ell & \ell_{2} & L \\
        0 & 0 & 0
    \end{pmatrix}^{2}
    r_{1} r_{2}\, g_{a}^{(L)} (r_{1}, r_{2}) B^{\rho_{1}}_{\rho_{0} \rho_{1}}
    \nonumber \\
    & \times \left[
    \rho_{\text{n}} (r_{1}) \rho_{\text{p}} (r_{1})
    + \rho_{\text{n}} (r_{2}) \rho_{\text{p}} (r_{2})
    \big] \big[
    \rho_{\text{n} \ell_{2} j_{2}} (r_{2}, r_{1})
    \delta_{\text{n} q}
    - \rho_{\text{p} \ell_{2} j_{2}} (r_{2}, r_{1})
    \delta_{\text{p} q}
    \right]\,\,.
    \label{eq:non-local-part-mean-field-rho1-spherical-symmetry}
\end{align}

%

Variation of~\eqref{eq:non-loc-s0-term-edf-spherical-symmetry} gives the two 
fields 
\begin{align}
     h_{q \ell j}^{\mathbf{s}_{0},\text{loc}} (r_{1})
    &= \frac{1}{32\pi^{2}}
    \sum_{\substack{L \ell_{1} j_{1}\\ \ell_{2} j_{2}}}
    \frac{(2L+1) (2\ell_{1}+1) (2\ell_{2}+1)}{
    \ell_{1} (\ell_{1}+1) \ell_{2} (\ell_{2}+1)}\,
    f(j_{1}, \ell_{1}, s) f(j_{2}, \ell_{2}, s) f(\ell_{1}, \ell_{2}, L)
    \nonumber \\
    & \times
        \begin{pmatrix}
        \ell_{1} & \ell_{2} & L \\
        0 & 0 & 0
    \end{pmatrix}^{2}
    \int\! \mathrm{d}r_{2}\,r_{2}^{2}\, g_{a}^{(L)} (r_{1}, r_{2})
    \nonumber \\
    &\times\left[
    \left( B^{\rho_{0}}_{\mathbf{s}_{0}\mathbf{s}_{0}}
    - B^{\rho_{0}}_{\mathbf{s}_{1}\mathbf{s}_{1}} \right)
    \rho_{0 \ell_{1} j_{1}} (r_{1}, r_{2})
    \rho_{0 \ell_{2} j_{2}} (r_{2}, r_{1})
    + 2 B^{\rho_{0}}_{\mathbf{s}_{1}\mathbf{s}_{1}} \sum_{q^{\prime}}
    \rho_{q^{\prime} \ell_{1} j_{1}} (r_{1}, r_{2})
    \rho_{q^{\prime} \ell_{2} j_{2}} (r_{2}, r_{1})
    \right]\,,
    \label{eq:local-part-mean-field-s0-spherical-symmetry}
\end{align}
and
\begin{align}
    h_{q \ell j}^{\mathbf{s}_{0},\text{nloc}} (r_{1}, r_{2})
    & = \frac{1}{8\pi} \sum_{L \ell_{2} j_{2}}
    \frac{(2L+1) (2\ell_{2}+1)}{\ell (\ell+1) \ell_{2} (\ell_{2}+1)}\,
    f(j, \ell, s) f(j_{2}, \ell_{2}, s) f(\ell, \ell_{2}, L)
    \begin{pmatrix}
        \ell & \ell_{2} & L \\
        0 & 0 & 0
    \end{pmatrix}^{2}
    r_{1} r_{2} 
    \nonumber \\
    & \times g_{a}^{(L)} (r_{1}, r_{2})\,\frac{\rho_{0} (r_{1}) + \rho_{0} (r_{2})}{2} \left[
    2 \big( B^{\rho_{0}}_{\mathbf{s}_{0}\mathbf{s}_{0}}
    - B^{\rho_{0}}_{\mathbf{s}_{1}\mathbf{s}_{1}} \big)
    \rho_{0 \ell_{2} j_{2}} (r_{1}, r_{2})
    + 4 B^{\rho_{0}}_{\mathbf{s}_{1}\mathbf{s}_{1}}
    \rho_{q \ell_{2} j_{2}} (r_{1}, r_{2})
    \right]\,.
    \label{eq:non-local-part-mean-field-s0-spherical-symmetry}
\end{align}
%
%
The variation of~\eqref{eq:non-loc-s1-term-edf-spherical-symmetry} yields
\begin{align}
     h^{\mathbf{s}_{1},\text{loc.}}_{q \ell j} (r_{1})
    &= \frac{1}{32\pi^{2}}
    \sum_{\substack{L \ell_{1} j_{1} \\ \ell_{2} j_{2}}}
    \frac{(2L+1) (2\ell_{1}+1) (2\ell_{2}+1)}{
    \ell_{1} (\ell_{1}+1) \ell_{2} (\ell_{2}+1)}\,
    f(j_{1}, \ell_{1}, s) f(j_{2}, \ell_{2}, s) f(\ell_{1}, \ell_{2}, L)
    \nonumber \\
    & \times
    \begin{pmatrix}
        \ell_{1} & \ell_{2} & L \\
        0 & 0 & 0
    \end{pmatrix}^{2}
    B^{\rho_{1}}_{\mathbf{s}_{0} \mathbf{s}_{1}}
    \int\! \mathrm{d}r_{2}\,r_{2}^{2}\,
    g_{a}^{(L)} (r_{1}, r_{2}) 
    \nonumber \\
    &\times
    \left[ 4 \rho_{q} (r_{1}) - 2 \rho_{0} (r_{1})
    \right] \left[
    \rho_{\text{n} \ell_{1} j_{1}} (r_{1}, r_{2})
    \rho_{\text{n} \ell_{2} j_{2}} (r_{2}, r_{1})
    - \rho_{\text{p} \ell_{1} j_{1}} (r_{1}, r_{2})
    \rho_{\text{p} \ell_{2} j_{2}} (r_{2}, r_{1})
    \right]\,,
    \label{eq:local-part-mean-field-s1-spherical-symmetry}
\end{align}
and
\begin{align}
    h^{\mathbf{s}_{1},\text{nloc}}_{q \ell j} (r_{1}, r_{2})
    & = -\frac{B^{\rho_{1}}_{\mathbf{s}_{0} \mathbf{s}_{1}}}{\pi} \sum_{L \ell_{2} j_{2}}
    \frac{(2L+1) (2\ell_{2}+1)}{\ell (\ell+1) \ell_{2} (\ell_{2}+1)}\,
    f(j, \ell, s) f(j_{2}, \ell_{2}, s) f(\ell, \ell_{2}, L)
    \begin{pmatrix}
        \ell & \ell_{2} & L \\
        0 & 0 & 0
    \end{pmatrix}^{2}
    r_{1} r_{2}
    \nonumber \\
    & \times g_{a}^{(L)} (r_{1}, r_{2}) \left[
    \rho_{\text{n}} (r_{1}) \rho_{\text{p}} (r_{1})
    + \rho_{\text{n}} (r_{2}) \rho_{\text{p}} (r_{2})
    \right] \left[
    \rho_{\text{n} \ell_{2} j_{2}} (r_{2}, r_{1})
    \delta_{\text{n} q}
    - \rho_{\text{p} \ell_{2} j_{2}} (r_{2}, r_{1})
    \delta_{\text{p} q}
    \right]\,.
    \label{eq:non-local-part-mean-field-s1-spherical-symmetry}
\end{align}
%
%
Finally, note that the pp
EDFs~\eqref{eq:non-loc-rhotq-term-edf-spherical-symmetry}
and~\eqref{eq:non-loc-stq-term-edf-spherical-symmetry} also give 
contributions to the normal field since they contain the density
$\rho_{\bar{q}}$. 
These contributions are local and read
\begin{align}
    h_{q \ell j}^{\tilde{\rho}_{q},\text{loc}} (r_{1})
    & = \frac{C^{\rho}_{\tilde{\rho}\tilde{\rho}}}{16\pi^{2}}
    \sum_{\substack{L \ell_{1} j_{1}\\ \ell_{2} j_{2}}}
    (2L+1) (2\ell_{1}+1) (2\ell_{2}+1)
    \begin{pmatrix}
        \ell_{1} & \ell_{2} & L \\
        0 & 0 & 0
    \end{pmatrix}^{2}
    \nonumber \\
    & \times \int\! \mathrm{d}r_{2}\,r_{2}^{2}\,
    g_{a}^{(L)} (r_{1}, r_{2})
    \tilde{\rho}_{\bar{q} \ell_{1}j_{1}} (r_{1}, r_{2})
    \tilde{\rho}_{\bar{q} \ell_{2}j_{2}} (r_{2}, r_{1})\,,
    \label{eq:local-part-mean-field-rhotq-spherical-symmetry}
\end{align}
and
\begin{align}
    h_{q \ell j}^{\tilde{\mathbf{s}}_{q},\text{loc}} (r_{1})
    & = \frac{C^{\rho}_{\tilde{\mathbf{s}}\tilde{\mathbf{s}}}}{32\pi^{2}}
    \sum_{\substack{L \ell_{1} j_{1}\\ \ell_{2} j_{2}}}
    \frac{(2L+1) (2\ell_{1}+1) (2\ell_{2}+1)}{
    \ell_{1} (\ell_{1}+1) \ell_{2} (\ell_{2}+1)}\,
    f(j_{1},\ell_{1},s) f(j_{2},\ell_{2},s) f(\ell_{1},\ell_{2},L)
    \begin{pmatrix}
        \ell_{1} & \ell_{2} & L \\
        0 & 0 & 0
    \end{pmatrix}^{2}
    \nonumber \\
    & \times \int\! \mathrm{d}r_{2}\,r_{2}^{2}\,
    g_{a}^{(L)} (r_{1}, r_{2})
    \tilde{\rho}_{\bar{q} \ell_{1}j_{1}} (r_{1}, r_{2})
    \tilde{\rho}_{\bar{q} \ell_{2}j_{2}} (r_{2}, r_{1})~.
    \label{eq:local-part-mean-field-stq-spherical-symmetry}
\end{align}


\subsubsection{Pairing mean field\label{subsubsec:pairing-mean-field-spherical-symmetry}}


Variation of~\eqref{eq:non-loc-rhotq-term-edf-spherical-symmetry} with 
respect to $\tilde{\rho}_{q \ell j}$ yields
\begin{align}
    \tilde{h}_{q \ell j}^{\tilde{\rho}_{q}, \text{nloc}} (r_{1}, r_{2})
    &= - \frac{C^{\rho}_{\tilde{\rho}\tilde{\rho}}}{4\pi}
    \sum_{L \ell_{1} j_{1}}
    (2L+1) (2\ell_{1}+1)
    \begin{pmatrix}
        \ell_{1} & \ell & L \\
        0 & 0 & 0
    \end{pmatrix}^{2} \nonumber \\
    &\hskip 1cm \times
    r_{1} r_{2}\, g_{a}^{(L)} (r_{1}, r_{2})
    \,\frac{\rho_{\bar{q}} (r_{1}) + \rho_{\bar{q}} (r_{2})}{2}\,
    \tilde{\rho}_{q\ell_{1}j_{1}} (r_{1}, r_{2})\,.
    \label{eq:non-local-part-pairing-mean-field-rhotq-spherical-symmetry}
\end{align}
%
%
The variation of~\eqref{eq:non-loc-stq-term-edf-spherical-symmetry} gives 
\begin{align}
    \tilde{h}_{q \ell j}^{\tilde{\mathbf{s}}_{q}, \text{nloc}} (r_{1}, r_{2})
    & = - \frac{C^{\rho}_{\tilde{\mathbf{s}}\tilde{\mathbf{s}}}}{8\pi}
    \sum_{L \ell_{1} j_{1}}
    \frac{(2L+1) (2\ell_{1}+1)}{\ell (\ell+1) \ell_{1} (\ell_{1}+1)}\,
    f(j_{1},\ell_{1},s) f(j,\ell,s) f(\ell_{1},\ell,L)
    \begin{pmatrix}
        \ell_{1} & \ell & L \\
        0 & 0 & 0
    \end{pmatrix}^{2}
    \nonumber \\
    & \hskip 1cm \times r_{1} r_{2}\, g_{a}^{(L)} (r_{1}, r_{2})
    \,\frac{\rho_{\bar{q}} (r_{1}) + \rho_{\bar{q}} (r_{2})}{2}\,
    \tilde{\rho}_{q\ell_{1}j_{1}} (r_{1}, r_{2})\,.
    \label{eq:non-local-part-pairing-mean-field-stq-spherical-symmetry}
\end{align}


\section{Conclusion\label{sec:conclusion}}

In this work, we have constructed the particle-hole and particle-particle 
parts of a nuclear EDF derived from a local leading-order three-body 
semi-regularised pseudopotential. 
Starting from the most general form of such pseudopotential, we carried out
the derivation of the corresponding contribution to the energy with the requirement that
it does not depend on local proton and neutron pairing densities and therefore 
does not require an energy cut-off when protons and neutrons are not mixed.

The contribution to the EDF that we 
obtain depends on two coupling constants that allow to separately
tune the strength of the interaction in the particle-hole and
particle-particle channels. This possibility was found
to be a missing feature in a previous attempt to build such a contribution to 
the EDF~\cite{costa_interactions_2022}.

We further specialised the derived functional to two limiting, but in practice most
relevant scenarios. First, analytical expressions for infinite nuclear matter
were obtained. This will enable direct evaluation of the functional's
behaviour in uniform systems.
Second, we derived the simplified form appropriate for spherically-symmetric
finite-nuclei which will support its forthcoming numerical implementation in a
coordinate-space solver~\cite{bennaceur_finres4_nodate} as well as in a code
used to adjust the parameters of the complete functional with two- and three-body terms.

The present work establishes the formal foundation necessary to implement such a
three-body semi-regularised pseudopotential in mean-field and also beyond-mean-field solvers.


\ack{We thank J.~Dobaczewski, M.~Kortelainen and D.~Lacroix for illuminating
  discussions on the construction of a three-body interaction. K.\,B.
  acknowledges the hospitality of the Physics Department of the University of
  Jyv\"askyl\"a where this work was initiated.}


\appendix 


\section{Coupling constants\label{app:coupling-constants}}

In the ph channel, the coupling constants of the local terms, expressed
using the parameters from~\eqref{eq:W1-as-function-of-W-beta-gamma} and~\eqref{eq:W2-as-function-of-W-beta-gamma}, are
\begin{alignat}{4}
  A^{\rho_{0}}_{\rho_{0} \rho_{0}}
  &=&\frac{1}{8}\, W_{3,1} &+& \frac{1}{16}\, W_{3,2}\,,
  \label{eq:cc-A-rho0-3b-1} \\
  A^{\rho_{0}}_{\rho_{1} \rho_{1}}
  &= -&\frac{1}{8}\, W_{3,1} &-&\frac{1}{16}\, W_{3,2}\,,
  \label{eq:cc-A-rho0-3b-2} \\
  A^{\rho_{0}}_{\mathbf{s}_{0} \mathbf{s}_{0}}
  &=&\frac{1}{24}\, W_{3,1} &+& \frac{1}{48}\, W_{3,2}\,,
  \label{eq:cc-A-rho0-3b-3} \\
  A^{\rho_{0}}_{\mathbf{s}_{1} \mathbf{s}_{1}}
  &= -&\frac{1}{24}\, W_{3,1} &-&\frac{1}{48}\, W_{3,2}\,,
  \label{eq:cc-A-rho0-3b-4} \\
  A^{\mathbf{s}_{0}}_{\rho_{0} \mathbf{s}_{0}}
  &=&&&\frac{1}{12}\, W_{3,2}\,,
  \label{eq:cc-A-rho0-3b-5} \\
  A^{\mathbf{s}_{0}}_{\rho_{1} \mathbf{s}_{1}}
  &= &&-& \frac{1}{12}\, W_{3,2}\,.
  \label{eq:cc-A-rho0-3b-6}
\end{alignat}
Those of the non-local terms are
\begin{alignat}{4}
  B^{\rho_{0}}_{\rho_{0} \rho_{0}}
  &= -&\frac{1}{16}\, W_{3,1} &-&\frac{3}{32}\, W_{3,2}\,,
  \label{eq:cc-B-rho0-3b-1} \\
  B^{\rho_{0}}_{\rho_{1} \rho_{1}}
  &= -&\frac{1}{16}\, W_{3,1} &-&\frac{3}{32}\, W_{3,2}\,,
  \label{eq:cc-B-rho0-3b-2} \\
  B^{\rho_{0}}_{\mathbf{s}_{0} \mathbf{s}_{0}}
  &= -&\frac{1}{16}\, W_{3,1} &-& \frac{1}{96}\, W_{3,2}\,,
  \label{eq:cc-B-rho0-3b-3} \\
  B^{\rho_{0}}_{\mathbf{s}_{1} \mathbf{s}_{1}}
  &= -&\frac{1}{16}\, W_{3,1} &-& \frac{1}{96}\, W_{3,2}\,,
  \label{eq:cc-B-rho0-3b-4} \\
  B^{\rho_{1}}_{\rho_{0} \rho_{1}}
  &=& \frac{1}{8}\, W_{3,1} &+& \frac{3}{16}\, W_{3,2}\,,
  \label{eq:cc-B-rho1-3b-1} \\
  B^{\rho_{1}}_{\mathbf{s}_{0} \mathbf{s}_{1}}
  &= &\frac{1}{8}\, W_{3,1} &+& \frac{1}{48}\, W_{3,2}\,, 
  \label{eq:cc-B-rho1-3b-2} \\
  B^{\mathbf{s}_{0}}_{\rho_{0} \mathbf{s}_{0}}
  &= - &\frac{1}{24}\, W_{3,1} &-& \frac{1}{16}\, W_{3,2}\,, 
  \label{eq:cc-B-s0-3b-1} \\
  B^{\mathbf{s}_{0}}_{\rho_{1} \mathbf{s}_{1}}
  &= - &\frac{1}{24}\, W_{3,1} &-& \frac{1}{16}\, W_{3,2}\,, 
  \label{eq:cc-B-s0-3b-2} \\
  B^{\mathbf{s}_{1}}_{\rho_{0} \mathbf{s}_{1}}
  &= &\frac{1}{24}\, W_{3,1} &+& \frac{1}{16}\, W_{3,2}\,,
  \label{eq:cc-B-s1-3b-1} \\
  B^{\mathbf{s}_{1}}_{\rho_{1} \mathbf{s}_{0}}
  &= &\frac{1}{24}\, W_{3,1} &+& \frac{1}{16}\, W_{3,2}\,, 
  \label{eq:cc-B-s1-3b-2}
\end{alignat}
and
\begin{alignat}{4}
    T^{\mathbf{s}_{0}}_{\mathbf{s}_{0} \mathbf{s}_{0}}
    &= - &\frac{1}{48}\, W_{3,1} &+& \frac{1}{96}\, W_{3,2}\,,
    \label{eq:cc-T-s0-s1-3b-1} \\
    T^{\mathbf{s}_{0}}_{\mathbf{s}_{1} \mathbf{s}_{1}}
    & = - &\frac{1}{48}\, W_{3,1} &+& \frac{1}{96}\, W_{3,2}\,, 
    \label{eq:cc-T-s0-s1-3b-2} \\
    T^{\mathbf{s}_{1}}_{\mathbf{s}_{0} \mathbf{s}_{1}}
    &=& \frac{1}{48}\, W_{3,1} &-& \frac{1}{96}\, W_{3,2}\,.
    \label{eq:cc-T-s0-s1-3b-3}
\end{alignat}
The coupling constants of the pp channel are 
\begin{alignat}{4}
    C^{\rho}_{\tilde{\rho} \tilde{\rho}}
    &= &\frac{1}{4}\, W_{3,1} &-& \frac{1}{8}\, W_{3,2}\,,
    \label{eq:cc-C-rho-3b-1} \\
    C^{\rho}_{\tilde{\mathbf{s}} \tilde{\mathbf{s}}}
    &= &\frac{1}{4}\, W_{3,1} &+& \frac{5}{24}\, W_{3,2}\,,
    \label{eq:cc-C-rho-3b-2} \\
    C^{\mathbf{s}}_{\tilde{\rho} \tilde{\mathbf{s}}}
    &= &\frac{1}{12}\, W_{3,1} &-& \frac{1}{24}\, W_{3,2}\,,
    \label{eq:cc-C-s-3b-1} \\
    C^{\mathbf{s}}_{\tilde{\mathbf{s}} \tilde{\rho}}
    &= &\frac{1}{12}\, W_{3,1} &-& \frac{1}{24}\, W_{3,2}\,,
    \label{eq:cc-C-s-3b-2} \\
    C^{\mathbf{s}}_{\tilde{\mathbf{s}} \tilde{\mathbf{s}}}
    &= &\frac{1}{12}\, W_{3,1} &+& \frac{1}{8}\, W_{3,2}\,.
    \label{eq:cc-C-s-3b-3} 
\end{alignat}
%


\section{Functional with proton-neutron mixing\label{app:functional-with-pn}}

Because of the isovector structure introduced by the breve representation,
additional notations can be defined for the operations between the densities
that are used to couple them to rank-0 tensors. 
We thus introduce the following notations for scalar products:
\begin{itemize}
    \item The scalar product between spin densities in space is, as usual, denoted
    with a dot "$\cdot$";
    \item The scalar product in isospin space is denoted with a circle
    "$\circ$";
    \item Combination of both scalar products is denoted "$\odot$".
\end{itemize}

The notations for cross products are:
\begin{itemize}
    \item The cross product between spin matrices in space is denoted "$\times$";
    \item The cross product in isospin space is denoted "$\wedge$";
    \item The combination of both cross products is denoted "$\diamond$".
\end{itemize}

With these notations, and using the parameters of the interaction
given by~\eqref{eq:W1-as-function-of-W-beta-gamma} and~\eqref{eq:W2-as-function-of-W-beta-gamma},
the particle-hole contribution to the EDF reads
\begin{align}
    E_{3,\text{ph}}
    & = \int\! \mathrm{d}^{3}r_{1}\,
    \mathrm{d}^{3}r_{2}\,g_{a} (\mathbf{r}_{12}) \Bigl\{
    \tfrac{1}{8} \left( W_{3,1} + \tfrac{1}{2}\, W_{3,2} \right)
    \rho_{0} (\mathbf{r}_{1})
    \left[ \rho_{0}^{2} (\mathbf{r}_{2}) - \pvec{\rho}^{2} (\mathbf{r}_{2})
    + \tfrac{1}{3}\, \mathbf{s}_{0}^{2} (\mathbf{r}_{2})
    - \tfrac{1}{3}\, \pvec{\mathbf{s}}^{2} (\mathbf{r}_{2})
    \right]
    \nonumber \\
    & + \tfrac{1}{12}\, W_{3,2} \left[
    \rho_{0} (\mathbf{r}_{1}) \mathbf{s}_{0} (\mathbf{r}_{1}) \cdot 
    \mathbf{s}_{0} (\mathbf{r}_{2})
    - \pvec{\rho} (\mathbf{r}_{1}) \circ \pvec{\mathbf{s}} (\mathbf{r}_{1})
    \cdot \mathbf{s}_{0} (\mathbf{r}_{2})
    \right]
    \nonumber \\
    & - \tfrac{1}{16}\, \rho_{0} (\mathbf{r}_{1})
    \left( W_{3,1} + \tfrac{3}{2}\, W_{3,2} \right) \left[
    \rho_{0} (\mathbf{r}_{1}, \mathbf{r}_{2}) 
    \rho_{0} (\mathbf{r}_{2}, \mathbf{r}_{1})
    + \pvec{\rho} (\mathbf{r}_{1}, \mathbf{r}_{2}) \circ 
    \pvec{\rho} (\mathbf{r}_{2}, \mathbf{r}_{1}) 
    \right]
    \nonumber \\
    & - \tfrac{1}{16}\, \rho_{0} (\mathbf{r}_{1})
    \left( W_{3,1} + \tfrac{1}{6}\, W_{3,2} \right) \left[
    \mathbf{s}_{0} (\mathbf{r}_{1}, \mathbf{r}_{2}) \cdot  
    \mathbf{s}_{0} (\mathbf{r}_{2}, \mathbf{r}_{1})
    + \pvec{\mathbf{s}} (\mathbf{r}_{1}, \mathbf{r}_{2}) \odot
    \pvec{\mathbf{s}} (\mathbf{r}_{2}, \mathbf{r}_{1}) 
    \right]
    \nonumber \\
    & + \tfrac{1}{16}\, \left[ \pvec{\rho} (\mathbf{r}_{1})
    + \pvec{\rho} (\mathbf{r}_{2})
    \right] \circ \left[
    \left( W_{3,1} + \tfrac{1}{6}\, W_{3,2} \right)
    \mathbf{s}_{0} (\mathbf{r}_{1}, \mathbf{r}_{2}) \cdot 
    \pvec{\mathbf{s}} (\mathbf{r}_{2}, \mathbf{r}_{1}) \right. \nonumber \\
    &\hskip 6cm\left.
    + \left( W_{3,1} + \tfrac{3}{2}\, W_{3,2} \right)
    \rho_{0} (\mathbf{r}_{1}, \mathbf{r}_{2}) 
    \pvec{\rho} (\mathbf{r}_{2}, \mathbf{r}_{1})
    \right]
    \nonumber \\
    & - \tfrac{1}{48}\, \left( W_{3,1} + \tfrac{3}{2}\, W_{3,2} \right) \left[
    \mathbf{s}_{0} (\mathbf{r}_{1})
    + \mathbf{s}_{0} (\mathbf{r}_{2})
    \right] \cdot \left[
    \rho_{0} (\mathbf{r}_{1}, \mathbf{r}_{2})
    \mathbf{s}_{0} (\mathbf{r}_{2}, \mathbf{r}_{1})
    + \pvec{\rho} (\mathbf{r}_{1}, \mathbf{r}_{2}) \circ 
    \pvec{\mathbf{s}} (\mathbf{r}_{2}, \mathbf{r}_{1})
    \right]
    \nonumber \\
    & + \tfrac{1}{48}\, \left( W_{3,1} + \tfrac{3}{2}\, W_{3,2} \right) \left[
    \pvec{\mathbf{s}} (\mathbf{r}_{1})
    + \pvec{\mathbf{s}} (\mathbf{r}_{2})
    \right] \odot \left[
    \rho_{0} (\mathbf{r}_{1}, \mathbf{r}_{2})
    \pvec{\mathbf{s}} (\mathbf{r}_{2}, \mathbf{r}_{1})
    + \pvec{\rho} (\mathbf{r}_{1}, \mathbf{r}_{2})
    \mathbf{s}_{0} (\mathbf{r}_{2}, \mathbf{r}_{1})
    \right]
    \nonumber \\
    & + \tfrac{\mathrm{i}}{16}\, \pvec{\rho} (\mathbf{r}_{1}) \circ \left[
    \left( W_{3,1} + \tfrac{1}{6}\, W_{3,2} \right)
    \pvec{\mathbf{s}} (\mathbf{r}_{1}, \mathbf{r}_{2}) \cdot \wedge \,
    \pvec{\mathbf{s}} (\mathbf{r}_{2}, \mathbf{r}_{1})
    + \left( W_{3,1} + \tfrac{3}{2}\, W_{3,2} \right)
    \pvec{\rho} (\mathbf{r}_{1}, \mathbf{r}_{2}) \wedge 
    \pvec{\rho} (\mathbf{r}_{2}, \mathbf{r}_{1})
    \right]
    \nonumber \\
    & - \tfrac{\mathrm{i}}{48} \left( W_{3,1} - \tfrac{1}{2}\, W_{3,2} \right)
    \mathbf{s}_{0} (\mathbf{r}_{1}) \cdot \left[
    \pvec{\mathbf{s}} (\mathbf{r}_{1}, \mathbf{r}_{2}) \times \circ\,
    \pvec{\mathbf{s}} (\mathbf{r}_{2}, \mathbf{r}_{1})
    + \mathbf{s}_{0} (\mathbf{r}_{1}, \mathbf{r}_{2}) \times 
    \mathbf{s}_{0} (\mathbf{r}_{2}, \mathbf{r}_{1})
    \right]
    \nonumber \\
    & + \tfrac{\mathrm{i}}{48} \left[
    \pvec{\mathbf{s}} (\mathbf{r}_{1})
    - \pvec{\mathbf{s}} (\mathbf{r}_{2})
    \right] \odot \left[
    \left( W_{3,1} + \tfrac{3}{2}\, W_{3,2} \right)
    \pvec{\mathbf{s}} (\mathbf{r}_{1}, \mathbf{r}_{2}) \wedge 
    \pvec{\rho} (\mathbf{r}_{2}, \mathbf{r}_{1}) \right. \nonumber \\
    &\hskip 6cm\left.
    + \left( W_{3,1} - \tfrac{1}{2}\, W_{3,2} \right)
    \pvec{\mathbf{s}} (\mathbf{r}_{1}, \mathbf{r}_{2}) \times
    \mathbf{s}_{0} (\mathbf{r}_{2}, \mathbf{r}_{1})
    \right]
    \nonumber \\
    & - \tfrac{1}{48} \left( W_{3,1} - \tfrac{1}{2}\, W_{3,2} \right)
    \pvec{\mathbf{s}} (\mathbf{r}_{1}) \odot\left[
    \pvec{\mathbf{s}} (\mathbf{r}_{1}, \mathbf{r}_{2}) \diamond \,
    \pvec{\mathbf{s}} (\mathbf{r}_{2}, \mathbf{r}_{1})\right]
    \Bigr\}\,,
    \label{eq:E-3b-ph-with-pn}
\end{align}
and the particle-particle contribution
\begin{align}
    E_{3,\text{pp}} = & \int\! \mathrm{d}^{3}r_{1}\,
    \mathrm{d}^{3}r_{2}\,g_{a} (\mathbf{r}_{12}) \Bigl\{
    \tfrac{1}{12} \left( W_{3,1} + \tfrac{1}{2}\, W_{3,2} \right)
    \rho_{0} (\mathbf{r}_{1}) \breve{\mathbf{s}}_{0} (\mathbf{r}_{2}) \cdot 
    \breve{\mathbf{s}}_{0}^{*} (\mathbf{r}_{2}) \nonumber \\
    &\hskip 4cm
    + \tfrac{\mathrm{i}}{24}\, W_{3,2}\, \mathbf{s}_{0} (\mathbf{r}_{1}) \cdot \left[
    \breve{\mathbf{s}}_{0} (\mathbf{r}_{2}) \times 
    \breve{\mathbf{s}}^{*}_{0} (\mathbf{r}_{2})
    \right]
    \nonumber \\
    & - \tfrac{1}{24} \left( W_{3,1} + \tfrac{3}{2}\, W_{3,2} \right)
    \breve{\mathbf{s}}_{0} (\mathbf{r}_{1}) \cdot \left[
    \breve{\mathbf{s}}_{0}^{*} (\mathbf{r}_{1}, \mathbf{r}_{2})
    \rho_{0} (\mathbf{r}_{2}, \mathbf{r}_{1})
    - \pvec{\breve{\mathbf{s}}}^{*} (\mathbf{r}_{1}, \mathbf{r}_{2}) \circ 
    \pvec{\rho} (\mathbf{r}_{2}, \mathbf{r}_{1})
    \right]
    \nonumber \\
    & + \tfrac{1}{24} \left( W_{3,1} - \tfrac{1}{2}\, W_{3,2} \right)
    \breve{\mathbf{s}}_{0} (\mathbf{r}_{1}) \cdot \left[
    \breve{\rho}_{0}^{*} (\mathbf{r}_{1}, \mathbf{r}_{2})
    \mathbf{s}_{0} (\mathbf{r}_{2}, \mathbf{r}_{1})
    - \pvec{\breve{\rho}}^{*} (\mathbf{r}_{1}, \mathbf{r}_{2}) \circ
    \pvec{\mathbf{s}} (\mathbf{r}_{2}, \mathbf{r}_{1})
    \right]
    \nonumber \\
    & -\tfrac{1}{12}\, W_{3,2}\,
    \breve{\mathbf{s}}_{0}^{*} (\mathbf{r}_{1}) \cdot \left[
    \breve{\mathbf{s}}_{0} (\mathbf{r}_{1}, \mathbf{r}_{2})
    \rho_{0} (\mathbf{r}_{1}, \mathbf{r}_{2})
    - \pvec{\breve{\mathbf{s}}} (\mathbf{r}_{1}, \mathbf{r}_{2}) \circ 
    \pvec{\rho} (\mathbf{r}_{1}, \mathbf{r}_{2})
    \right]
    \nonumber \\
    & + \tfrac{1}{16} \left( W_{3,1} - \tfrac{1}{2}\, W_{3,2} \right)
    \rho_{0} (\mathbf{r}_{1}) \left[
    \breve{\rho}_{0}^{*} (\mathbf{r}_{1}, \mathbf{r}_{2})
    \breve{\rho}_{0} (\mathbf{r}_{1}, \mathbf{r}_{2})
    + \pvec{\breve{\rho}}^{*} (\mathbf{r}_{1}, \mathbf{r}_{2}) \circ 
    \pvec{\breve{\rho}} (\mathbf{r}_{1}, \mathbf{r}_{2})
    \right]
    \nonumber \\
    & + \tfrac{1}{16} \left( W_{3,1} + \tfrac{5}{6}\, W_{3,2} \right)
    \rho_{0} (\mathbf{r}_{1}) \left[
    \breve{\mathbf{s}}_{0}^{*} (\mathbf{r}_{1}, \mathbf{r}_{2}) \cdot 
    \breve{\mathbf{s}}_{0} (\mathbf{r}_{1}, \mathbf{r}_{2})
    + \pvec{\breve{\mathbf{s}}}^{*} (\mathbf{r}_{1}, \mathbf{r}_{2}) \odot 
    \pvec{\breve{\mathbf{s}}} (\mathbf{r}_{1}, \mathbf{r}_{2})
    \right]
    \nonumber \\
    & - \tfrac{1}{16} \left( W_{3,1} - \tfrac{1}{2}\, W_{3,2} \right)
    \pvec{\rho} (\mathbf{r}_{1}) \circ \left[
    \pvec{\breve{\rho}}^{*} (\mathbf{r}_{1}, \mathbf{r}_{2})
    \breve{\rho}_{0} (\mathbf{r}_{1}, \mathbf{r}_{2})
    + \breve{\rho}_{0}^{*} (\mathbf{r}_{1}, \mathbf{r}_{2})
    \pvec{\breve{\rho}} (\mathbf{r}_{1}, \mathbf{r}_{2})
    \right]
    \nonumber \\
    & - \tfrac{1}{16} \left( W_{3,1} + \tfrac{5}{6}\, W_{3,2} \right)
    \pvec{\rho} (\mathbf{r}_{1}) \circ \left[
    \pvec{\breve{\mathbf{s}}}^{*} (\mathbf{r}_{1}, \mathbf{r}_{2}) \cdot
    \breve{\mathbf{s}}_{0} (\mathbf{r}_{1}, \mathbf{r}_{2})
    + \breve{\mathbf{s}}_{0}^{*} (\mathbf{r}_{1}, \mathbf{r}_{2}) \cdot 
    \pvec{\breve{\mathbf{s}}} (\mathbf{r}_{1}, \mathbf{r}_{2})
    \right]
    \nonumber \\
    & + \tfrac{1}{48} \left( W_{3,1} - \tfrac{1}{2}\, W_{3,2} \right)
    \mathbf{s}_{0} (\mathbf{r}_{1}) \cdot \left[
    \breve{\rho}_{0}^{*} (\mathbf{r}_{1}, \mathbf{r}_{2})
    \breve{\mathbf{s}}_{0} (\mathbf{r}_{1}, \mathbf{r}_{2})
    + \breve{\mathbf{s}}_{0}^{*} (\mathbf{r}_{1}, \mathbf{r}_{2})
    \breve{\rho}_{0} (\mathbf{r}_{1}, \mathbf{r}_{2})
    \right.\nonumber \\
    & \hspace{4cm}
    + \left.\pvec{\breve{\rho}}^{*} (\mathbf{r}_{1}, \mathbf{r}_{2}) \circ
    \pvec{\breve{\mathbf{s}}} (\mathbf{r}_{1}, \mathbf{r}_{2})
    + \pvec{\breve{\mathbf{s}}}^{*} (\mathbf{r}_{1}, \mathbf{r}_{2}) \circ
    \pvec{\breve{\rho}} (\mathbf{r}_{1}, \mathbf{r}_{2})
    \right]
    \nonumber \\
    & - \tfrac{1}{48} \left( W_{3,1} - \tfrac{1}{2}\, W_{3,2} \right)
    \pvec{\mathbf{s}} (\mathbf{r}_{1}) \odot \left[
    \breve{\rho}_{0}^{*} (\mathbf{r}_{1}, \mathbf{r}_{2})
    \pvec{\breve{\mathbf{s}}} (\mathbf{r}_{1}, \mathbf{r}_{2})
    + \pvec{\breve{\mathbf{s}}}^{*} (\mathbf{r}_{1}, \mathbf{r}_{2})
    \breve{\rho}_{0} (\mathbf{r}_{1}, \mathbf{r}_{2})
    \right.\nonumber \\
    & \hspace{4cm}
    + \left.\pvec{\breve{\rho}}^{*} (\mathbf{r}_{1}, \mathbf{r}_{2})
    \breve{\mathbf{s}}_{0} (\mathbf{r}_{1}, \mathbf{r}_{2})
    + \breve{\mathbf{s}}_{0}^{*} (\mathbf{r}_{1}, \mathbf{r}_{2})
    \pvec{\breve{\rho}} (\mathbf{r}_{1}, \mathbf{r}_{2})
    \right]
    \nonumber \\
    & - \tfrac{\mathrm{i}}{16} \pvec{\rho} (\mathbf{r}_{1}) \circ \left[
    \left( W_{3,1} + \tfrac{5}{6}\, W_{3,2} \right)
    \pvec{\breve{\mathbf{s}}} (\mathbf{r}_{1}, \mathbf{r}_{2}) \cdot \wedge \,
    \pvec{\breve{\mathbf{s}}}^{*} (\mathbf{r}_{1}, \mathbf{r}_{2}) \right. \nonumber \\
    &\hskip 4cm\left.
    + \left( W_{3,1} - \tfrac{1}{2}\, W_{3,2} \right)
    \pvec{\breve{\rho}} (\mathbf{r}_{1}, \mathbf{r}_{2}) \wedge
    \pvec{\breve{\rho}}^{*} (\mathbf{r}_{1}, \mathbf{r}_{2})
    \right]
    \nonumber \\
    & + \tfrac{\mathrm{i}}{48} \left( W_{3,1} + \tfrac{3}{2}\, W_{3,2} \right)
    \mathbf{s}_{0} (\mathbf{r}_{1}) \cdot \left[
    \pvec{\breve{\mathbf{s}}} (\mathbf{r}_{1}, \mathbf{r}_{2}) \times \circ \, 
    \pvec{\breve{\mathbf{s}}}^{*} (\mathbf{r}_{1}, \mathbf{r}_{2})
    + \breve{\mathbf{s}}_{0} (\mathbf{r}_{1}, \mathbf{r}_{2}) \times 
    \breve{\mathbf{s}}_{0}^{*} (\mathbf{r}_{1}, \mathbf{r}_{2})
    \right]
    \nonumber \\
    & - \tfrac{\mathrm{i}}{48} \left( W_{3,1} + \tfrac{3}{2}\, W_{3,2} \right)
    \pvec{\mathbf{s}} (\mathbf{r}_{1}) \odot \left[
    \breve{\mathbf{s}}_{0} (\mathbf{r}_{1}, \mathbf{r}_{2}) \times 
    \pvec{\breve{\mathbf{s}}}^{*} (\mathbf{r}_{1}, \mathbf{r}_{2})
    + \pvec{\breve{\mathbf{s}}} (\mathbf{r}_{1}, \mathbf{r}_{2}) \times 
    \breve{\mathbf{s}}_{0}^{*} (\mathbf{r}_{1}, \mathbf{r}_{2})
    \right]
    \nonumber \\
    & - \tfrac{\mathrm{i}}{48} \left( W_{3,1} - \tfrac{1}{2}\, W_{3,2} \right)
    \pvec{\mathbf{s}} (\mathbf{r}_{1}) \odot \left[
    \pvec{\breve{\mathbf{s}}} (\mathbf{r}_{1}, \mathbf{r}_{2}) \wedge 
    \pvec{\breve{\rho}}^{*} (\mathbf{r}_{1}, \mathbf{r}_{2})
    + \pvec{\breve{\rho}} (\mathbf{r}_{1}, \mathbf{r}_{2}) \wedge 
    \pvec{\breve{\mathbf{s}}}^{*} (\mathbf{r}_{1}, \mathbf{r}_{2})
    \right]
    \nonumber \\
    & + \tfrac{1}{48} \left( W_{3,1} + \tfrac{3}{2}\, W_{3,2} \right)
    \pvec{\mathbf{s}} (\mathbf{r}_{1}) \odot \left[
    \pvec{\breve{\mathbf{s}}} (\mathbf{r}_{1}, \mathbf{r}_{2}) \diamond \,
    \pvec{\breve{\mathbf{s}}}^{*} (\mathbf{r}_{1}, \mathbf{r}_{2})
    \right]
    \nonumber \\
    & - \tfrac{\mathrm{i}}{24} \left( W_{3,1} + \tfrac{1}{2}\, W_{3,2} \right)
    \breve{\mathbf{s}}_{0} (\mathbf{r}_{1}) \cdot \left[
    \pvec{\mathbf{s}} (\mathbf{r}_{2}, \mathbf{r}_{1}) \times \circ \,
    \pvec{\breve{\mathbf{s}}}^{*} (\mathbf{r}_{1}, \mathbf{r}_{2})
    - \mathbf{s}_{0} (\mathbf{r}_{2}, \mathbf{r}_{1}) \times 
    \breve{\mathbf{s}}_{0}^{*} (\mathbf{r}_{1}, \mathbf{r}_{2})
    \right]
    \nonumber \\
    & + \tfrac{\mathrm{i}}{24} \left( W_{3,1} + \tfrac{1}{2}\, W_{3,2} \right)
    \breve{\mathbf{s}}_{0}^{*} (\mathbf{r}_{1}) \cdot \left[
    \pvec{\mathbf{s}} (\mathbf{r}_{1}, \mathbf{r}_{2}) \times \circ \,
    \pvec{\breve{\mathbf{s}}} (\mathbf{r}_{1}, \mathbf{r}_{2})
    - \mathbf{s}_{0} (\mathbf{r}_{1}, \mathbf{r}_{2}) \times 
    \breve{\mathbf{s}}_{0} (\mathbf{r}_{1}, \mathbf{r}_{2})
    \right]
    \Bigr\}\,.
    \label{eq:E-3b-pp-with-pn}
\end{align}


\section{Auxiliary functions\label{app:auxiliary-functions}}

The auxiliary functions $F_{0}$, $G_{0}$, $H_{0}$ and $K_{0}$ are given by
\begin{align}
    F_{0} (x) &= \frac{12}{x^{3}} \left[ 
    \frac{1-\mathrm{e}^{-x^{2}}}{x^{3}}
    - \frac{3-\mathrm{e}^{-x^{2}}}{2x}
    + \frac{\sqrt{\pi}}{2} \Erf x
    \right]\,,    \label{eq:auxiliary-function-F0} \\
    G_{0} (x)& = \frac{12}{x^{6}} \left( \mathrm{e}^{-x^{2}} - 1 \right)
    + \frac{6}{x^{4}} \left( \mathrm{e}^{-x^{2}} + 1 \right)\,,
    \label{eq:auxiliary-function-G0} \\
     H_{0} (x)& = \frac{1}{x^{2}} \left(
    1 - \mathrm{e}^{-x^{2}}\right)\,,
    \label{eq:auxiliary-function-H0} \\
     K_{0} (x) &= \frac{6}{x^{4}} \left(
    \mathrm{e}^{-x^{2}} - 1 \right)
    +  \frac{3\sqrt{\pi}}{x^{3}}\, \Erf x\,.
    \label{eq:auxiliary-function-K0}
\end{align}












\bibliography{bibliography}
\bibliographystyle{iopart-num}

\end{document}